\documentclass[aps,preprint]{revtex4}
\usepackage{amsfonts}
\usepackage{amsmath}
\usepackage{amssymb}
\usepackage{graphicx}
\usepackage{subfigure}
\usepackage{mathrsfs, mathtools}
\usepackage{pdfpages}
\usepackage{multirow}
\usepackage{float}
\usepackage{color}

\usepackage{doi}
\usepackage{hyperref}
\hypersetup{
  colorlinks=true,        
  linkcolor=blue,         
  citecolor=cyan,         
}

\usepackage{xcolor}
\usepackage{orcidlink}
\usepackage{caption}
\usepackage{lineno,hyperref}
\hypersetup{colorlinks =true, allcolors = blue}
\providecommand{\U}[1]{\protect\rule{.1in}{.1in}}

\begin{document}

\begin{center}
{\Large \textbf{Schwarzschild Black Hole in Galaxies Surrounded by a 
Dark Matter Halo}}
\end{center}

\begin{center}

Ahmad Al-Badawi \orcidlink{0000-0002-3127-3453}\\
Department of Physics, Al-Hussein Bin Talal University, P. O. Box: 20, 71111,
Ma'an, Jordan. 
\bigskip E-mail: ahmadbadawi@ahu.edu.jo\\
Sanjar Shaymatov\orcidlink{0000-0002-5229-7657}\\
Institute for Theoretical Physics and Cosmology, Zhejiang University of Technology, Hangzhou 310023, China\\
Institute of Fundamental and Applied Research, National Research University TIIAME, Kori Niyoziy 39, Tashkent 100000, Uzbekistan\\
University of Tashkent for Applied Sciences, Str. Gavhar 1, Tashkent 100149, Uzbekistan\\
Western Caspian University, Baku AZ1001, Azerbaijan\\
E-mail: sanjar@astrin.uz\\
Yassine Sekhmani\orcidlink{0000-0001-7448-4579}\\
Center for Theoretical Physics, Khazar University, 41 Mehseti Street, Baku, AZ1096, Azerbaijan.\\
Centre for Research Impact and Outcome, Chitkara University Institute of Engineering and Technology, Chitkara University, Rajpura, 140401, Punjab, India\\
Chitkara Centre for Research and Development, Chitkara University, Baddi, Himachal Pradesh, 174103, India\\
E-mail: sekhmaniyassine@gmail.com

{\Large Abstract} \end{center}

In this paper, we derive a novel Schwarzschild-like black hole (BH) solution describing a static and asymptotically flat BH surrounded by a dark matter (DM) halo with a Dehnen-type density distribution in the surrounding environment. We investigate the properties of the obtained BH by studying the curvature properties and energy conditions in Einstein gravity. Furthermore, we explore the features of a novel Schwarzschild-like BH embedded in a DM halo with Dehnen-type density profile by analyzing the timelike geodesics of particles along with BH observable properties.

\section{Introduction}

Our universe contains massive objects such as BHs that play crucial roles in Einstein's general relativity and other gravity theories, making the idea that they are completely isolated in our universe implausible. BHs probable live in dynamic and complex surroundings. In example, there is considerable evidence that supermassive BHs are the driving forces underlying active galactic nuclei \cite{Rees84ARAA,Kormendy95ARAA}. There is also strong evidence that DM surrounds most galaxies in a halo \cite{Bertone18Nature}. Furthermore, 
 astronomical evidence for DM can be found in  
 measurements of galactic rotation curves \cite{Rubin70ApJ}, bullet clusters \cite{Corbelli00MNRAS}, baryon acoustic oscillations and cosmic microwave background \cite{Komatsu11ApJS} and large-scale structure development of the universe \cite{Davis85ApJ}. These observations are unlikely to 
 have satisfactory interpretations unless we assume a large amount of DM in our universe. The cosmic microwave background observation shows that the 
 Universe is composed mostly of DM (27\%) and dark energy (68\%). Several unknown particles predicted by theories beyond the Standard Model are 
 potential candidates for DM \cite{Boehm04NPB,Bertone05PhR,Feng09JCAP,Schumann19}, including weakly interacting massive particles (WIMPs), axions, and sterile neutrinos. A halo of DM surrounds most galaxies as well \cite{Bertone18Nature}, which could have significant effects on supermassive BHs. The investigation of DM halo effects on supermassive BHs is therefore of great importance, as it allows us to gain a better 
 understanding of how DM interacts with BHs. There are many effects of DM halo structures on galactic rotation curves \cite{Rubin70ApJ,Bertone18Nature,Corbelli00MNRAS,Sofue2020RotationCO}, matter motion observed in bullet cluster collisions \cite{Clowe06ApJL}, BH shadows and polarized images \cite{Jusufi20EPJC,Das_2022} and  gravitational lensing of massive objects \cite{Karamazov21ApJ,Qi23}. 
 
 It is worth noting that, despite the fact that DM is only known for its gravitational interaction feature, there is strong evidence confirming the presence of DM in the universe (see, for example, \cite{Bertone05,deSwart17Nat,Wechsler18}). It is a well-established fact through observations that stars can be formed in particular regions of galaxies, i.e., very close to their centers. This results in a galactic DM halo being formed in the surrounding regions of host galaxies. Hence, host galaxies (i.e., spiral or giant elliptical galaxies) are formed in the presence of a DM halo (see, for example, \cite{Valluri04ApJ,Akiyama19L1,Akiyama19L6,Akiyama22L12}). Analyzing DM models can help enhance our understanding of its fundamental nature, despite being a long-standing problem  GR. Based on observational data, the DM halo only helps explain the rotational velocity of stars orbiting host galaxies \cite{Persic96}. Many different DM halo models have been developed based on simulation studies and astrophysical data. Various BH solutions have since been developed, such as those involving a DM profile associated with a phantom scalar field \cite{Li-Yang12}, explored with extensive analyses providing some insights into the nature of DM profile \cite{Hendi20,Rizwan19,Shaymatov21d,Rayimbaev-Shaymatov21a,Shaymatov21pdu,Shaymatov22a} and analytical models that exhibit a supermassive BH solution immersed in a DM halo (see, e.g., \cite{Cardoso22DM,Hou18-dm,Shen24PLB}). Besides, the Navarro-Frenk-White  model \cite{Navarro96ApJ}, the Einasto model \cite{Dutton14MNRAS,Merritt06ApJ}, the Burkert model \cite{Burkert95ApJL}, and Dehnen model \cite{Dehnen93}.  
 Recently, a lot of studies into the Dehnen-type DM halo effects on BHs have been conducted from various perspectives. In Ref. \cite{Shukirgaliyev21A&A}, the authors 
 investigate the impact of density profile slopes on the survivability of low  star-formation efficiency 
 star clusters following immediate gas expulsion. To accomplish this goal, they examine scenarios in which a star cluster has a Plummer profile and a Dehnen 
 profile with cusps of differing slopes at the time of formation. Other BH solution with the Dehnen DM halo profile is constructed in \cite{Pantig22JCAP} for the study of an ultrafaint 
 dwarf galaxy. Further, in \cite{Gohain24DM} a new BH solution is surrounded by a Dehnen-(1,4,0) type DM halo created by embedding a Schwarzschild black hole within the halo, resulting in a composite DM-BH system. This study \cite{Gohain24DM} examines the thermodynamics of the effective BH spacetime and its null geodesic. 
 
 In this paper, we present a new Schwarzschild-like BH solution describing a static and asymptotically flat BH surrounded by a DM halo with a Dehnen-type density profile $(1, 4, 5/2)$ in its surroundings. To achieve this aim, we employ the method developed in Ref.~\cite{Matos05}, which has also been used by the authors of Ref.~\cite{Xu18JCAP}. We examine the singularity structure of this BH at $r = 0$, focusing on analyzing the spacetime curvature invariants and then consider the energy conditions in Einstein gravity. Finally, we delve into how the DM profile affects particles' timelike geodesics, enhancing our understanding of the effects of the DM halo profile on particle geodesics.
 
 The paper is organized in a following way: In Sec.~\ref{sec2},  we obtain the metric function of the BH-DM profile by examining the theoretical framework of the Dehnen-type density distribution. Sec.~\ref{sec3} examines the singularity structure and then the energy conditions in Einstein gravity. In Sec.~\ref{sec4}, we investigate the timelike geodesics of massive particles with the rest mass around the Schwarzschild-like BH immersed in the DM halo within a Dehnen-type density profile.  Finally, we end up with our remarks and conclusions in Sec.~\ref{sec5}.

\section{Schwarzschild-like BH metric immersed in the DM halo} \label{sec2}

In this section,  we utilize the method developed in Ref.~\cite{Matos05} and later employed by Ref.~\cite{Xu18JCAP} to determine the Schwarzschild-like BH metric in the DM halo within a Dehnen-type density distribution. This method typically consists of two steps: First, in the case of general relativity, the DM spacetime metric is constructed using the DM profile; second, the approximate solution for the BH under DM is derived by analyzing Einstein's field equations. The first step towards a Schwarzschild BH is to understand the mass distribution of a Dehnen-type DM halo. In a spherically symmetric spacetime, the mass distribution is determined by the DM density profile. The mass profile is given by \cite{Mo10book}: 
 \begin{equation}
M_{D}=4\pi \int\limits_{0}^{r}\rho \left( r^{\prime }\right) r^{\prime
2}dr^{\prime },
\end{equation}
 where the density profile of the Dehnen type DM
halo is given by 
 \begin{equation}
\rho =\rho _{s}\left( \frac{r}{r_{s}}\right) ^{-\gamma }\left[ \left( 
\frac{r}{r_{s}}\right) ^{\alpha }+1\right] ^{\frac{\gamma -\beta }{\alpha }},
\label{dens1}
\end{equation}%
where, $\rho _{s}$ and $r_{s}$ are  the central halo density radius.  Whereas  $\gamma$ determines the specific variant of the profile. The values of $\gamma$ lies within
$[0, 3]$ i.e. $\gamma = 3/2$ is used to fit the surface brightness profiles of elliptical galaxies which closely
resembles the de Vaucouleurs $r^{1/4}$ profile \cite{Shakeshaft74}. In this work, we use the Dehnen-$\left( \alpha
,\beta ,\gamma \right) =\left( 1,4,5/2\right)$ DM halo. Hence Eq. (\ref{dens1})
becomes 
\begin{equation}
\rho =\frac{\rho _{s}}{\left( \frac{r}{r_{s}}\right) ^{5/2}\left( \frac{r}{r_{s}}+1\right) ^{3/2}}.
\label{dens2}
\end{equation}
Inserting Eq. (\ref{dens2}) in the mass profile 
 relation we obtain 
\begin{equation}
M_{D}=4\pi \int\limits_{0}^{r}\frac{\rho _{s}r^{\prime
2}}{\left( \frac{r^{\prime}}{r_{s}}\right) ^{5/2}\left( \frac{r^{\prime}}{r_{s}}+1\right) ^{3/2}} dr^{\prime }=\frac{8\pi \rho _{s}r_{s}^{3}}{\sqrt{1+\frac{r_s}{r}}}.
\end{equation}
In a spherically symmetric spacetime, the mass distribution at the center of the DM halo can be used to calculate the tangential velocity of the particle moving within it as follow:
\begin{equation}
v_{D}^{2}=\frac{M_{D}}{r}=\frac{8\pi \rho _{s}r_{s}^{3}}{r\sqrt{1+\frac{r_s}{r}}}.
\end{equation}
A pure DM halo can be described by a spherically symmetric line element:
\begin{equation}
ds^{2}=-A\left( r\right) dt^{2}+\frac{dr^{2}}{B\left( r\right) }+r^{2}\left(
d\theta ^{2}+\sin ^{2}\theta d\phi ^{2}\right),\label{m2}
\end{equation}%
where $A(r)$ represents the redshift functions, and $B(r)$ represents the shape functions. It is important to highlight that this BH metric (\ref{m2}) comply with Newtonian approximate, thus $A(r) = B(r)$. The metric function $A(r)$ is related to the tangential velocity through the
relation 
\begin{equation}
v_{D}^{2}=r\frac{d}{dr}\left( \ln\sqrt{ A(r)}\right) .
\end{equation}%
Solving for $A(r)$ we obtain 
\begin{equation}
A(r)=exp\left[- 32\pi \rho _{s}r_{s}^{3}\sqrt{\frac{r+r_s}{r_s^2\,r}}\right] \approx 1-32\pi \rho _{s}r_{s}^{3}\sqrt{\frac{r+r_s}{r_s^2\,r}},
\end{equation}
where a  leading order term was retained in the equation. 

\begin{figure}
    \centering
    \includegraphics[scale=0.85]{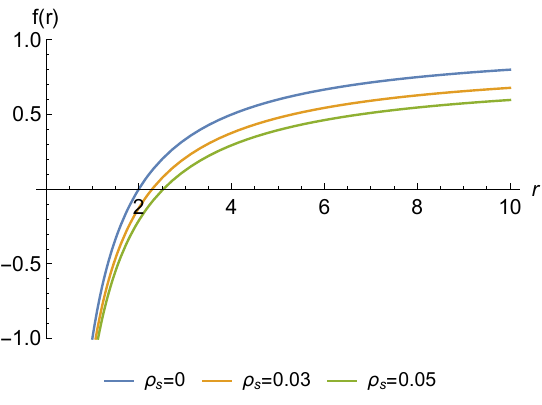}\includegraphics[scale=0.85]{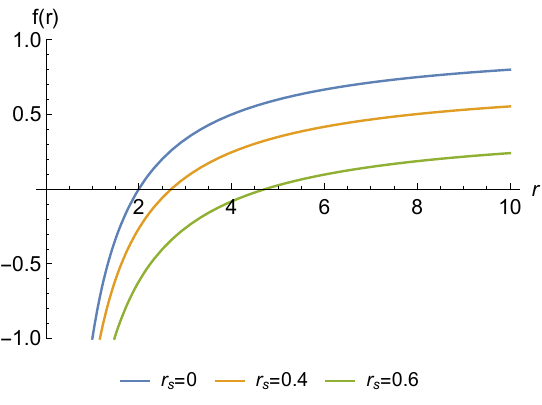}
    \caption{Lapse function $f(r)$ for varying values of $\rho_s$ by setting $r_s=0.2$ (left) and for varying values of $r_s$ by setting $\rho_s=0.015$ (right)}
    \label{figa1}
\end{figure}

The second step follows Xu's method \cite{Xu18JCAP},  which has also been used by others \cite{Azreg-Ainou14PRD,Jusufi20EPJC,Hou18JCAP,Gohain24DM,Xu21JCAP,Yang24EPJC}, is to solve Einstein's field
equation for a Dehnen-type DM halo distribution and a point mass
distribution. When the Dehnen-type DM halo profile completely determines the energy momentum tensor $T_{\mu \nu }$, Einstein's field equation follows.: 
\begin{equation}
R_{\mu \nu }-\frac{1}{2}Rg_{\mu \nu }=\kappa ^{2}T_{\mu \nu }(D)
\end{equation}%
where $\kappa ^{2}=8\pi ,$\ $R_{\mu \nu }$ denotes the Ricci tensor, $R$ the
Ricci scalar $g_{\mu \nu }$ is  the metric tensor. In addition, $T_{\mu
\nu }(D)$ denotes the energy-momentum tensor of the Dehnen type DM halo
spacetime, which can be written  as $T_{\mu }^{\nu }=g^{\nu \alpha }T_{\mu
\alpha }=diag[-\rho ,p_{r},p,p].$ Thus, the Einstein's field equation become%
\begin{equation}
\kappa ^{2}T_{t}^{t}(D)=B(r)\left( \frac{1}{r}\frac{B^{\prime }(r)}{B(r)}+%
\frac{1}{r^{2}}\right) -\frac{1}{r^{2}},
\end{equation}%
\begin{equation}
\kappa ^{2}T_{r}^{r}(D)=B(r)\left( \frac{1}{r}\frac{A^{\prime }(r)}{A(r)}+%
\frac{1}{r^{2}}\right) -\frac{1}{r^{2}},
\end{equation}%
\begin{equation}
\kappa ^{2}T_{\theta }^{\theta }(D)=\frac{1}{%
2}B(r)\left( \frac{A^{\prime \prime }(r)A(r)-A^{\prime 2}(r)}{A^{2}(r)}+%
\frac{A^{\prime 2}(r)}{2A^{2}(r)}+\frac{1}{2}\left( \frac{%
A^{\prime }(r)}{A(r)}+\frac{B^{\prime }(r)}{B(r)}\right) +\frac{A^{\prime
}(r)B^{\prime }(r)}{2A(r)B(r)}\right) .
\end{equation}%
\begin{equation}
   \kappa ^{2}T_{\phi }^{\phi }(D)=\kappa ^{2}T_{\theta }^{\theta }(D)
\end{equation}
Following \cite{Xu18JCAP}, for the combined system of the Schwarzschild BH and the DM
halo, the spacetime metric can be written as 
\begin{equation}
ds^{2}=-\left( A(r)+F_{1}(r)\right) dt^{2}+\frac{1}{B(r)+F_{2}(r)}%
dr^{2}+r^{2}\left( d\theta ^{2}+\sin ^{2}\theta d\phi ^{2}\right) ,
\label{feid2}
\end{equation}%
where the functions $F_{1}(r)$ and $F_{2}(r)$ are determined by BH
parameters and DM halo parameters. Therefore, the Einstein field equation
can now be written as%
\begin{equation}
R_{\mu \nu }-\frac{1}{2}Rg_{\mu \nu }=\kappa ^{2}\left[ T_{\mu \nu
}(D)+T_{\mu \nu }(BH)\right] ,  \label{fied1}
\end{equation}%
where $T_{\mu \nu }(BH)$ arises from the matter content of the pure BH
spacetime.  Using the combined spacetime metric (\ref{feid2}), then the field
equation (\ref{fied1}) can be simplified as 
\begin{equation}
\left( B(r)+F_{2}(r)\right) \left( \frac{1}{r}\frac{B^{\prime
}(r)+F_{2}^{\prime }(r)}{B(r)+F_{2}(r)}+\frac{1}{r^{2}}\right) =B(r)\left( 
\frac{1}{r}\frac{B^{\prime }(r)}{B(r)}+\frac{1}{r^{2}}\right) ,
\end{equation}%
\begin{equation}
\left( B(r)+F_{2}(r)\right) \left( \frac{1}{r}\frac{A^{\prime
}(r)+F_{1}^{\prime }(r)}{A(r)+F_{1}(r)}+\frac{1}{r^{2}}\right) =B(r)\left( 
\frac{1}{r}\frac{A^{\prime }(r)}{A(r)}+\frac{1}{r^{2}}\right) ,
\end{equation}%
Taking Schwarzschild BH as the boundary condition, then, one obtains the
solutions to the above differential equations as
\begin{equation}
F_{2}(r)=\frac{-2M}{r}.
\end{equation}%
\begin{equation}
F_{1}(r)=\exp \left[ \int \frac{B(r)}{B(r)-\frac{2M}{r}}\left( \frac{1}{r}+
\frac{A^{\prime }(r)}{A(r)}\right)dr \right] -A(r),
\end{equation}%
Based on the assumption of $A(r)=B(r)$, we find that $F_{1}(r)=F_{2}(r)=%
\frac{-2M}{r}.$ Therefore, the  Schwarzschild BH in the Dehnen-$(1,4,5/2)$  DM halo
can be written as 
\begin{equation} 
ds^{2}=-f\left( r\right) dt^{2}+\frac{dr^{2}}{f\left( r\right) }+r^{2}\left(
d\theta ^{2}+\sin ^{2}\theta d\phi ^{2}\right) ,\label{m1}
\end{equation} where
\begin{equation}
f\left( r\right) =1-\frac{2M}{r}-32\pi \rho _{s}r_{s}^{3}\sqrt{\frac{r+r_s}{r_s^2\,r}}\, .  \label{laps1}
\end{equation}
Fig.~\ref{figa1} depicts how parameters $\rho_s$ and $r_s$ affect the lapse function $f(r)$ for various values. As both parameter values grow, so does the horizon. This plot clearly shows how core density and core radius of the DM halo play a significant role in BH horizons' existence. 
The horizons of the Schwarzschild-like BH surrounded by a Dehnen type DM halo can be found using the constraint $f(r) = 0$. Each combination of the DM halo parameters $\rho_s$ and $r_s$ results in a unique horizon which is given by \begin{equation}
    r_h=\frac{2 M+512 \pi^2 r_s^5 \rho_s^2+32\sqrt{2} \pi r_s^2 \rho_s\sqrt{2M^2 +Mr_s+128 \pi^2 r_s^6 \rho_s^2}}{1-1024 \pi^2 r_s^4 \rho_s^2}. \label{h1}
\end{equation}
\begin{figure}
    \centering
    \includegraphics[scale=0.65]{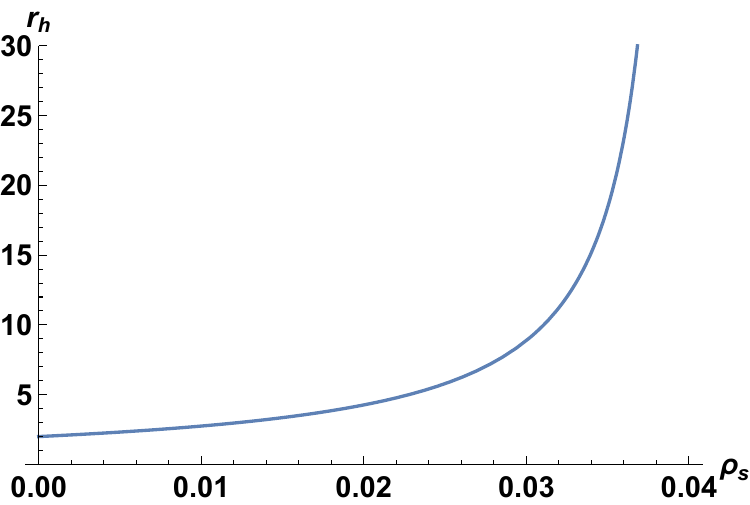}\includegraphics[scale=0.65]{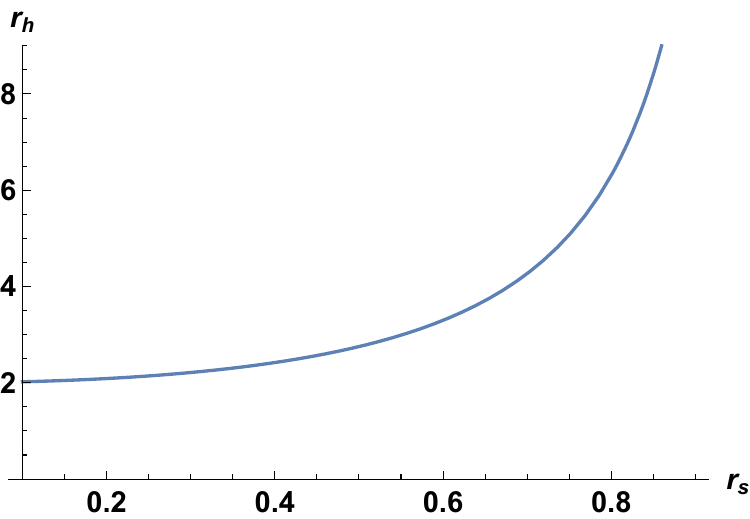}
    \caption{Variation of horizon $r_h$ for varying values of $\rho_s$ by setting $r_s=0.5$ (left) and for varying values of $r_s$ by setting $\rho_s=0.01$ (right)}
    \label{figa2}
\end{figure}
Fig.~\ref{figa2} shows how the horizon depends on the parameters $\rho_s$ and $r_s$, indicating that the horizon increases with these parameters.

\section{Curvature properties and Energy conditions}\label{sec3}

To examine the singularity structure of this BH (\ref{m1}) at $r = 0$, it becomes crucial to analyse the curvature invariants of the spacetime such as Ricci scalar, $R$, Ricci square, $R_{\mu\nu}R_{\mu\nu}$, and Kretshmann scalars, $R_{\mu\nu\alpha\beta}R_{\mu\nu\alpha\beta}$. 
\begin{equation}
R=\frac{8\pi r_{s}^{2}\rho _{s}\left( 8r^{2}+12rr_{s}+3r_{s}^{2}\right) }{%
r^{4}\left( \frac{r+r_{s}}{r}\right) ^{3/2}},\label{r5}
\end{equation}
\begin{eqnarray}
R_{\mu\nu}R_{\mu\nu} &=& \frac{32\pi ^{2}r_{s}^{4}\left(
64r^{4}+192r^{3}r_{s}+208r^{2}r_{s}^{2}+96rr_{s}^{3}+17r_{s}^{4}\right) \rho
_{s}^{2}}{r^{5}\left( r+r_{s}\right) ^{3}}\, ,\label{rr}\\
R^{\mu \nu \alpha \beta }R_{\mu \nu \alpha \beta }&=&\frac{16}{r^{8}}\left[
r^{2}\left( M+\frac{8\pi r_{s}^{3}\rho _{s}}{\sqrt{\frac{r+r_{s}}{r}}}%
\right) ^{2}+r^{2}\left( 16\pi rr_{s}^{2}\rho _{s}\sqrt{\frac{r+r_{s}}{r}}%
\right) ^{2}\right.\nonumber\\&&+\left. \left( Mr+\frac{2\pi r_{s}^{3}\rho _{s}\left( 4r+3r_{s}\right) 
}{\left( \frac{r+r_{s}}{r}\right) ^{3/2}}\right) ^{2}\right]. \label{rrr}
\end{eqnarray}
\begin{figure}
\begin{center}
\includegraphics[scale=0.55]{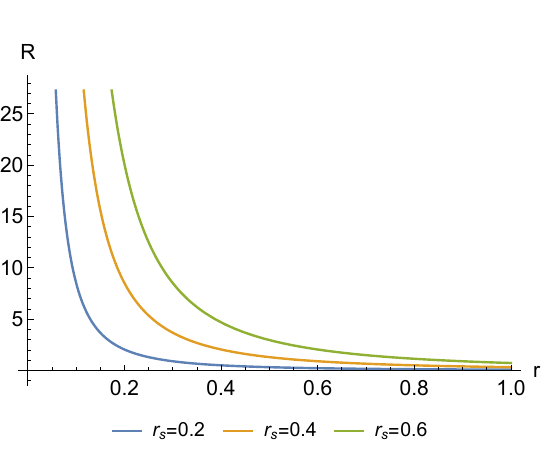}
\includegraphics[scale=0.55]{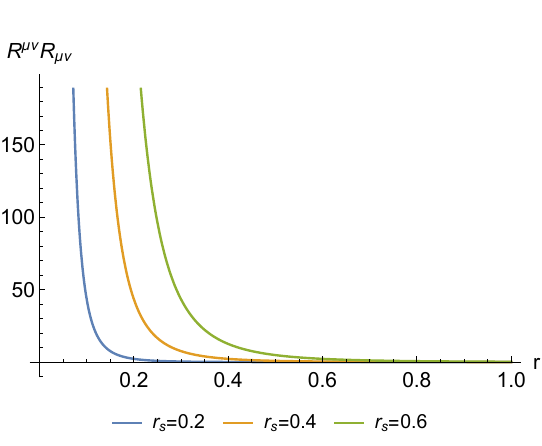}
\includegraphics[scale=0.55]{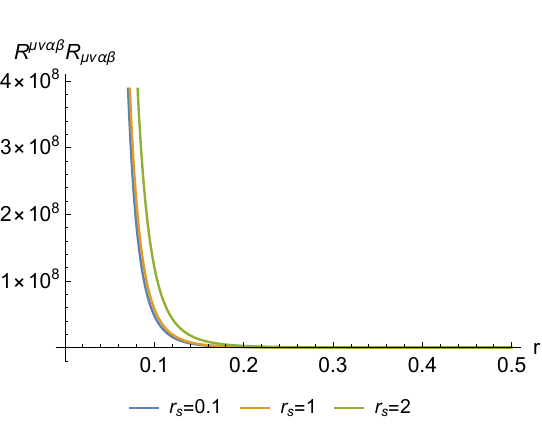}
\end{center}
\caption{Variation of Ricci, Ricci squared, and Kretschmann scalar vs $r$ for different values of core radius. Here ($M=1$ and $\rho_s=0.01$}\label{figa3}
\end{figure}

On closer inspection, it appears that the Schwarzschild-like BH solution immersed in Dehnen type DM sets to be singular at the center $( r=0)$ based on the relevant analysis of the invariants (\ref{r5}), (\ref{rr}), and (\ref{rrr}). 
It is therefore possible to obtain the results by inspecting the behavior at the center,
\begin{equation}
   \lim\limits_{r\to 0} \,
    R,\, R_{\mu\nu}R^{\mu\nu}, \,
R_{\mu\nu\alpha\beta}R^{\mu\nu\alpha\beta}\approx\infty.
    \end{equation}
On the other hand, the inspection at large distance shows that
\begin{equation}
   \lim\limits_{r\to \infty}R,
     \vspace{3mm}\\ R_{\mu\nu}R^{\mu\nu},
     \vspace{3mm}\\
R_{\mu\nu\alpha\beta}R^{\mu\nu\alpha\beta}=0
\end{equation}
It states that at large distances, the Ricci scalar, Ricci square, and Kretschmann scalar all have a defined finite term (Fig. \ref{figa3}). Essentially, these invariants show that our BH solution represented by the metric (\ref{m1}) is unique and that there is a physical singularity at $r = 0$.
\begin{figure*}[!htp]
      	\centering{
       \includegraphics[scale=0.54]{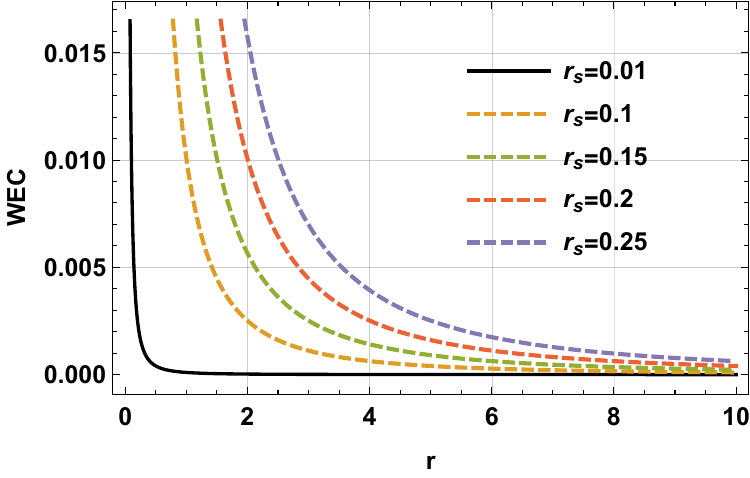} \hspace{2mm}
      	\includegraphics[scale=0.55]{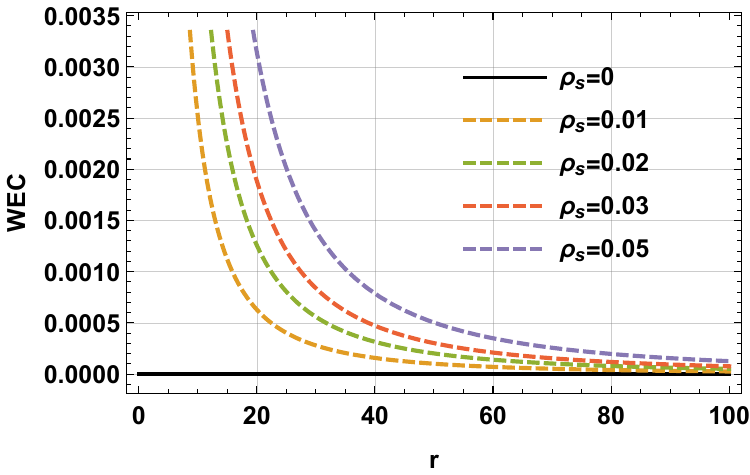} \hspace{2mm}
      	}
      	\centering{
       \includegraphics[scale=0.54]{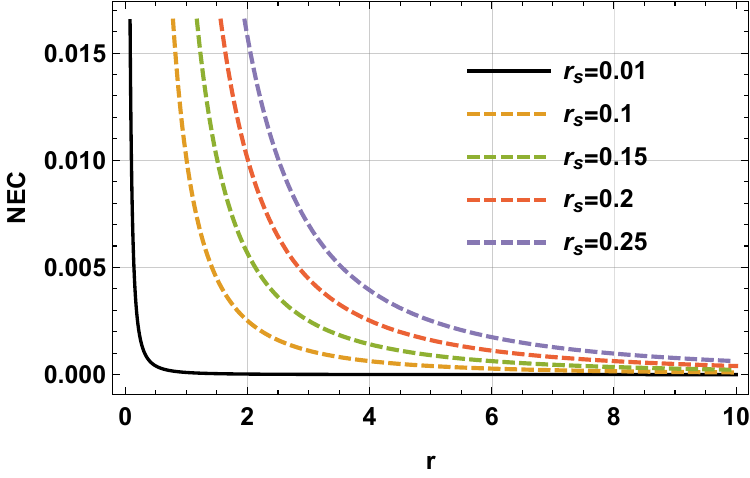} \hspace{2mm}
      	\includegraphics[scale=0.55]{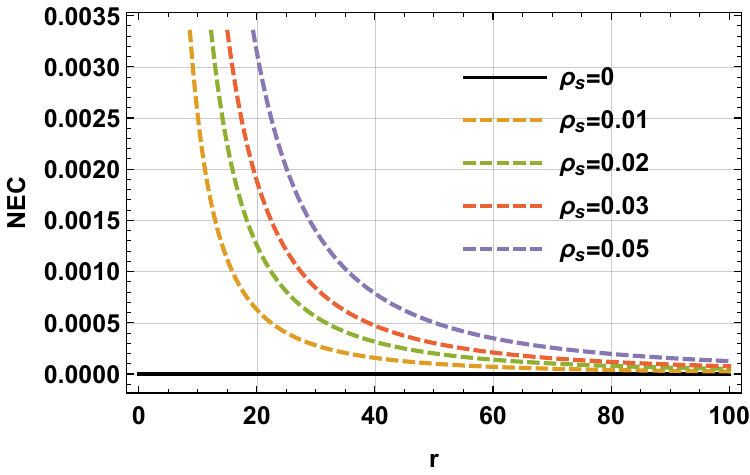} \hspace{2mm}
      }

      	\centering{
       \includegraphics[scale=0.54]{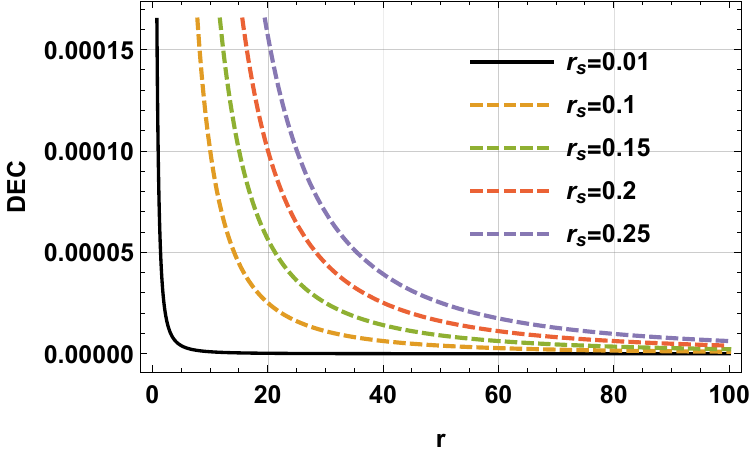} \hspace{2mm}
       \includegraphics[scale=0.55]{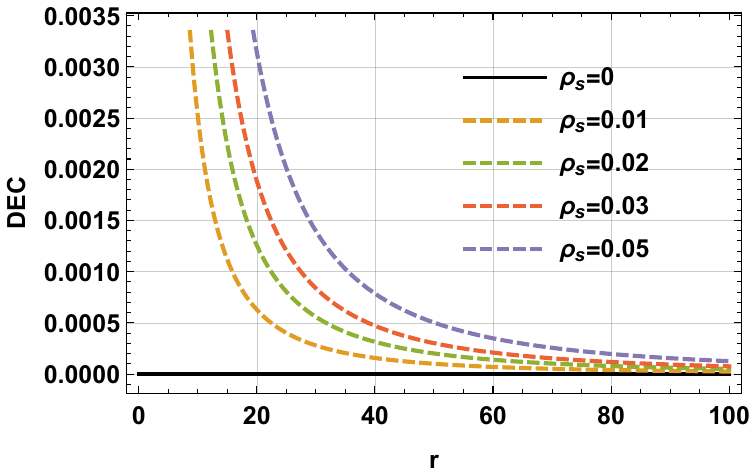} \hspace{2mm}
      	}

      	\centering{
       \includegraphics[scale=0.58]{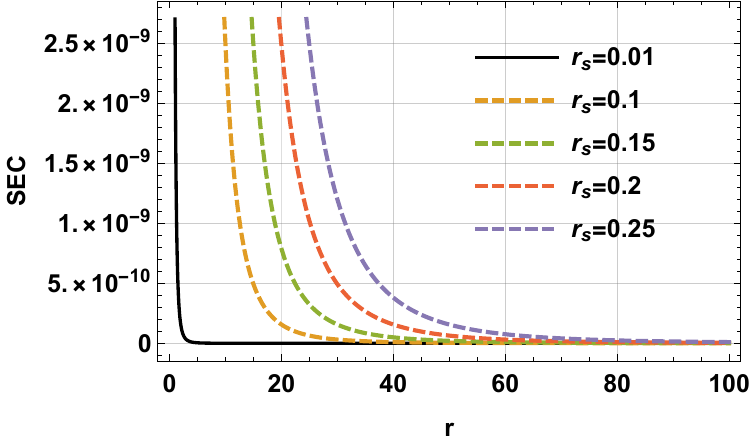} \hspace{2mm}
       \includegraphics[scale=0.59]{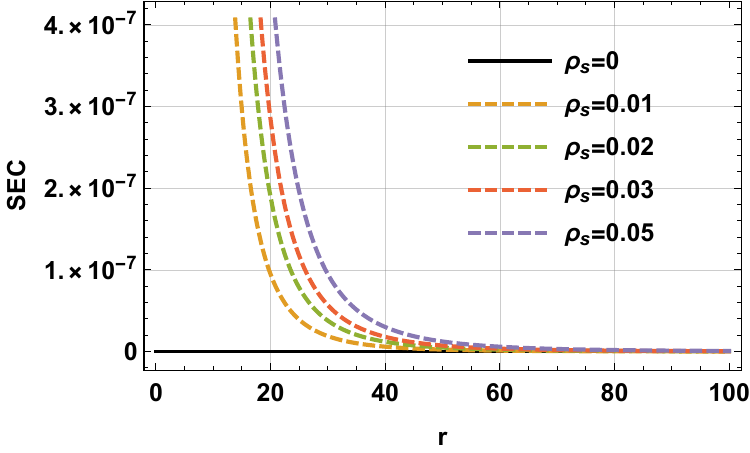} \hspace{2mm}
      	}
      	\caption{The variation of $\rho$ (weak energy condition), $\rho+\sum_i P_i$ (strong energy condition), $\rho+P_{\theta,\phi} $ (null energy condition), and $\rho-\mid P_{\theta,\phi}\mid $ (dominant energy condition) versus $r$ for multiple values of the parameter $r_s$ with $\rho_s=0.01$ (left panel) and for multiple value of $\rho_s$ with $r_s=0.5$ (right panel).}
      	\label{ECs}
      \end{figure*}

Our next step is to investigate the energy conditions in Einstein gravity. In view of supplying appropriate physical constraints on the gravitational configuration with matter field contribution, we probe the constraints of energy conditions within the resulting BH spacetime~\cite{Hawking:1973uf, Ghosh:2008zza,Kothawala:2004fy, Ghosh:2020syx}.

The stress-energy tensor elements $T_{\mu\nu}$ ruled by the Einstein equations \eqref{fied1} for the Dehnen-type DM halo BHs consist of the following:
        \begin{align}
        \rho&=-P_r=\frac{16 \pi  r_s \left(r_s+2 r\right) \rho _s}{r^3 \sqrt{\frac{r_s+r}{r r_s^2}}},\label{rho}\\
      P_\theta&=P_\phi=\frac{4 \pi  r_s \rho _s}{r^4 \left(\frac{r_s+r}{r r_s^2}\right){}^{3/2}}.\label{P}
    \end{align}
    By tracking these components, we can proceed to inspect the energy condition constraints to deliver the particular feature in query of the BH solution.
    \begin{itemize}
        \item The weak energy condition (WEC) holds that $T_{\mu\nu}\, t^\mu t^\nu\geqslant0$ anywhere, regardless of the time vector $t^\mu$, which is roughly equated to~\cite{Toshmatov:2017kmw}
    \begin{equation}
        \rho\ge0,\quad \rho+ P_i\ge0\quad (i=r, \theta, \phi)
    \end{equation}
    with $\rho+P_r=0$, so
    \begin{align}
        &\rho+P_\theta=\frac{4 \pi  \left(8 r^2+12 r r_s+5 r_s^2\right) \rho _s}{r^4 r_s \left(\frac{r_s+r}{r r_s^2}\right)^{3/2}}.
   \label{3-}
    \end{align}

The WEC is satisfied if the requirements set out here are fulfilled
\begin{equation}
   \frac{16 \pi  r_s \left(r_s+2 r\right) \rho _s}{r^3 \sqrt{\frac{r_s+r}{r r_s^2}}}\geqslant0,
\end{equation}
which demonstrates the perfect satisfaction of the WEC.

\item The null energy condition (NEC) implies that $T_{\mu\nu}\, t^\mu t^\nu\geqslant0$ within global space-time for all null vectors $t^\mu$. The NEC demands $\rho+P_r \geqslant 0$ which is precisely zero, and $\rho+P_\theta=\rho+P_\phi \geqslant 0$, which is satisfied for the equation \eqref{3-} whenever the condition is satisfied that
\begin{equation}
   \frac{4 \pi  \left(8 r^2+12 r r_s+5 r_s^2\right) \rho _s}{r^4 r_s \left(\frac{r_s+r}{r r_s^2}\right){}^{3/2}}\geqslant0,
\end{equation}
predicting full satisfaction of the NEC constraint.
\item The dominant energy condition (DEC) states that, in tandem with WEC, whatever the future-directed causal vector field ( either temporal or zero) $\mathbf{Y}$, $- T^\mu_\nu\mathbf{Y}^\nu$ has to be a future-directed causal vector. In other words, mass-energy can apparently always be found to be moving faster than light. On the other hand, DEC requires that $\rho-\mid P_{\theta,\phi}\mid\geqslant0$ provides the inequality
\begin{equation}
  \frac{4 \pi  \left(8 r^2+12 r r_s+3 r_s^2\right) \rho _s}{r^4 r_s \left(\frac{r_s+r}{r r_s^2}\right){}^{3/2}}\geqslant0,
\end{equation}
which ensures complete fulfillment of this energy condition.
\item The strong energy condition (SEC) asserts that $T_{\mu\nu}, t^\mu t^\nu\geqslant 1/2 \,T_{\mu\nu} t^\nu t_\nu$ overall, just for every time vector $t^\mu$ which means that~\cite{Toshmatov:2017kmw}
\begin{equation}
        \rho+\sum_i P_i=\frac{8 \pi  r_s \rho _s}{r^4 \left(\frac{r_s+r}{r r_s^2}\right){}^{3/2}}\ge0,
        \label{37}
    \end{equation}
    so that, the SEC is also satisfied.
    \end{itemize}

  The analytical examination of the energy conditions gives intriguing results, allowing us to presume that all the energy conditions are satisfied. The graphical aspect, meanwhile, proves this characteristic for the BHs of the Dehnen-type DM halo. To highlight this behaviour, Fig. \ref{ECs} offers a suitable graphical analysis that presents the variation of WEC, NEC, DEC, and SEC against the radial spacetime $r$.

\section{Timelike geodesics}\label{sec4}

Hereafter, we consider timelike geodesics of massive particles with the rest mass $m$ around the Schwarzschild BH surrounded by the Dehnen-$(1,4,5/2)$ type DM halo. To this end we first write the Hamiltonian for the given system as follows~\cite{Misner73}:  
\begin{eqnarray}
 H  \equiv  \frac{1}{2}& g^{\alpha\beta}&\frac{\partial \mathcal{S}}{\partial
x^{\alpha}}\frac{\partial \mathcal{S}}{\partial
x^{\beta}}\, ,
\label{Eq:H}
\end{eqnarray}
where $S$ and $x^{\alpha}$ represents the action and the four-vector coordinate, respectively. We then turn to the Hamiltonian of the given system, which is defined by $H=k/2$ with $k/m^2=-1$. The action $S$ then reads as 
\begin{eqnarray}\label{Eq:sep1}
\mathcal{S}= -\frac{1}{2}k\lambda-Et+L\varphi+\mathcal{S}_{r}(r)+\mathcal{S}_{\theta}(\theta)\ ,
\end{eqnarray}
where $S_{r}$ and $S_{\theta}$ are interpreted as the functions of $r$ and $\theta$, respectively. Eq.~(\ref{Eq:sep1}) allows one to rewrite the Hamilton-Jacobi equation as \cite{Shaymatov21pdu,Shaymatov22a}  
\begin{eqnarray}\label{Eq:sep}
&-&
\frac{E^2}{f(r)}
+f(r)\left(\frac{\partial \mathcal{S}_{r}}{\partial
r}\right)^{2}+{1\over r^{2}}\left(\frac{\partial \mathcal{S}_{\theta}}{\partial
\theta}\right)^{2}+\frac{L^{2}}{r^{2} \sin^2\theta }-k=0 \, . 
\end{eqnarray}
In the separated form of the Hamilton-Jacobi equation, the four quantities, such as specific energy $E$, angular momentum $L$, and $k$~\cite{Misner73}, occur as the conserved quantities of the motion, while the fourth one can be ignored  when the latitudinal motion takes place at the equatorial plane, $\theta=\pi/2$. 
\begin{figure*}
\begin{center}
\begin{tabular}{c c}
  \includegraphics[scale=0.465]{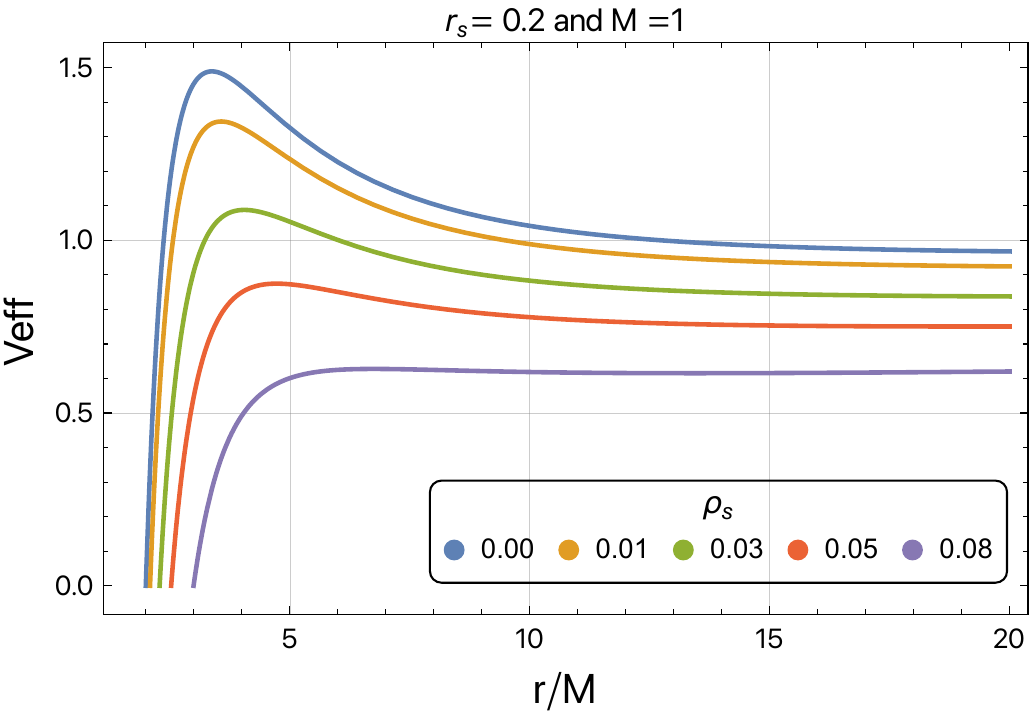}\hspace{0cm}
  \includegraphics[scale=0.465]{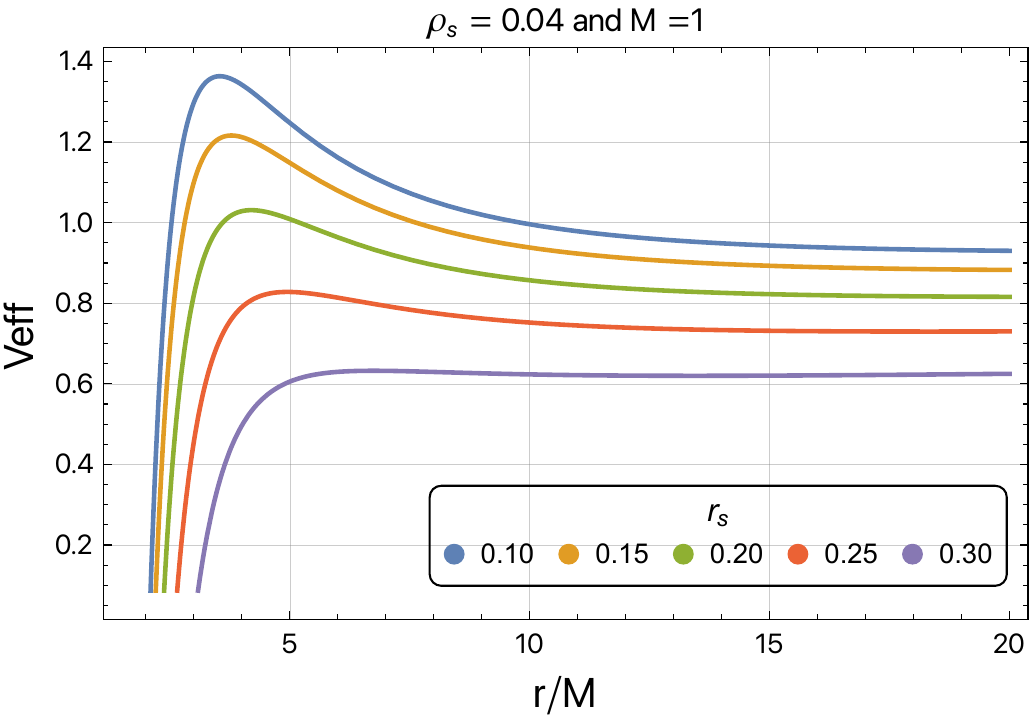}\hspace{0cm}
\end{tabular}
	\caption{\label{fig:eff_pot} 
The effective potential for the timelike motion of particles orbiting around the Schwarzschild BH surrounded by the Dehnen-type DM halo. Left panel: $V_{eff}$ is plotted for different values of the DM density profile parameter $\rho_s$ while keeping $r_m$ fixed. Right panel: $V_{eff}$ is plotted for different values of the halo core radius parameter $r_s$ while keeping $\rho_s$ fixed.}
\end{center}
\end{figure*}
\begin{figure*}
\begin{tabular}{cccc}
  \includegraphics[scale=0.45]{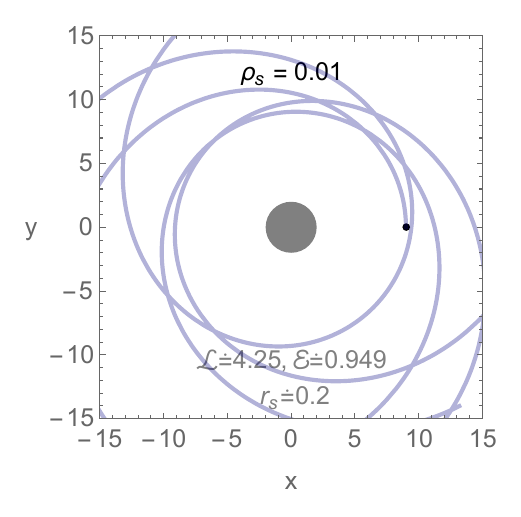}\hspace{-0.2cm}
  \includegraphics[scale=0.45]{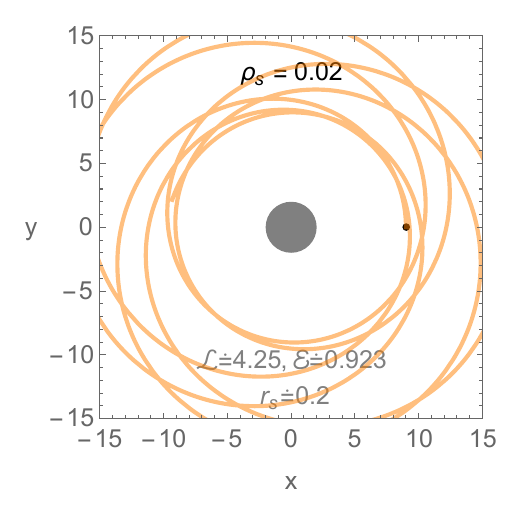}\hspace{-0.2cm}
  \includegraphics[scale=0.45]{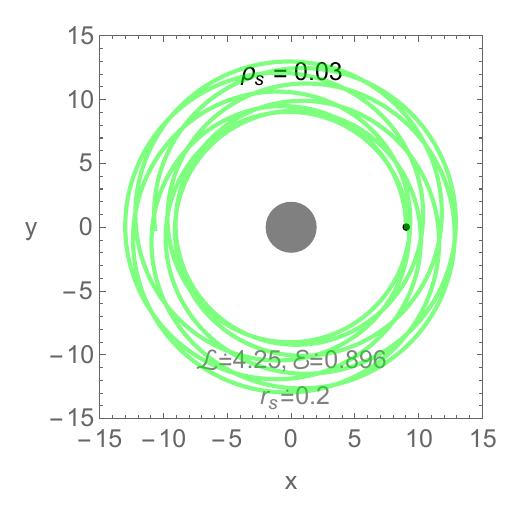}\hspace{-0.2cm}
\includegraphics[scale=0.45]{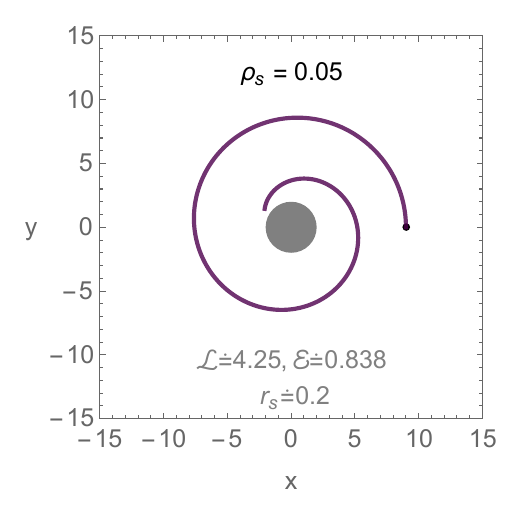}\\

  \includegraphics[scale=0.45]{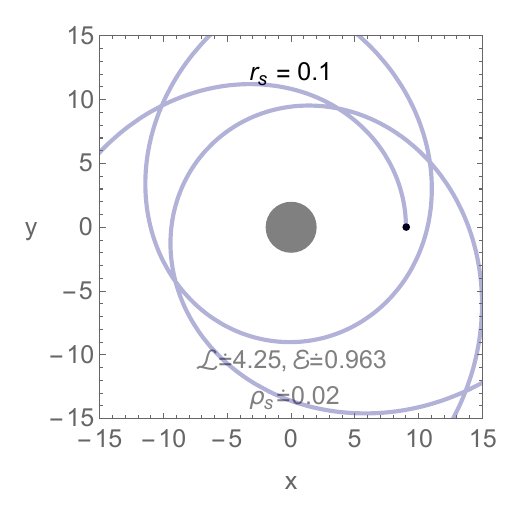}\hspace{-0.2cm}
  \includegraphics[scale=0.45]{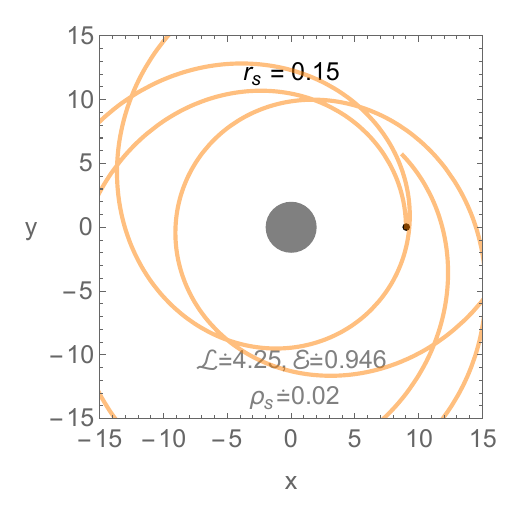}\hspace{-0.2cm}
  \includegraphics[scale=0.45]{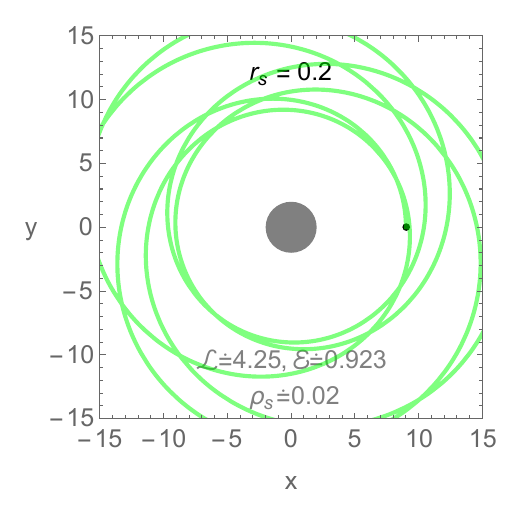}\hspace{-0.2cm}
  \includegraphics[scale=0.45]{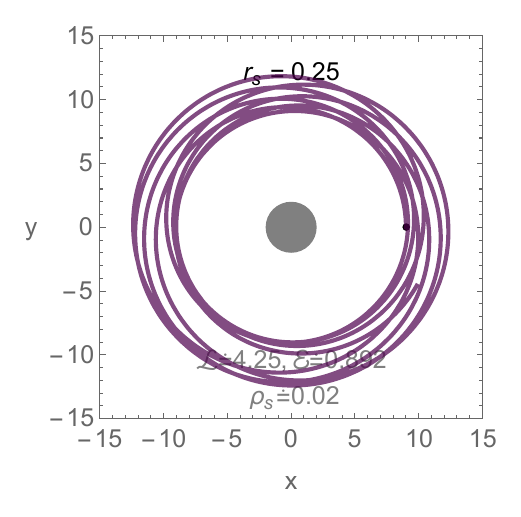}
  \end{tabular}
\caption{\label{fig:timelike} The geodesics of timelike particles orbiting around the Schwarzschild BH surrounded by the Dehnen-type DM halo for various combinations of the DM halo parameters $\rho_s$ (top row) and $r_s$ (bottom row). Note that the behavior of timelike particle geodesics are observed from the polar view, i.e. $z = 0$.}
\end{figure*}
Additionally, separable functions $S_{r}$ and $S_{\theta}$ in the Hamiltonian equation can be defined by 
\begin{align}
&S_r=\int\sqrt{E^2-f(r)\left(-k+\frac{K}{r^2}\right)}\,\frac{dr}{f(r)}\ , \label{HJ1} \\ &S_\theta=\int \sqrt{K-\frac{L^2}{\sin^2\theta}}\,d\theta\ , \label{HJ2}
\end{align}
where $K$ represents the well-known Carter constant. For making analysis simple, we shall further use 
\begin{eqnarray}
{\cal E}&=&\frac{E}{m}\ ,~ {\cal L}=\frac{L}{mM}\ ,~ {\cal K}=\frac{K}{(mM)^2}\, \mbox{\,\,\,and\,\,\,}\frac{k}{m^2}=-1\, .
\end{eqnarray} 
Following to the Hamiltonian $m\, dx^\alpha/d\lambda=g^{\alpha\beta}\partial S/\partial x^\beta$ with an affine parameter $\lambda$, we define the equations of motion as follows:  
\begin{eqnarray}\label{Eq:tdot}
\dot{t}&=&\frac{{\cal E}}{f(r)}\ ,\\\dot{\phi}&=&\frac{{\cal L}}{r^2\sin^2\theta}\ ,\\
\label{Eq:rdot}\dot{r}^2&=&{\cal E}^2-f(r)\left(1+\frac{\cal K}{r^2}\right)
\geq 0\ , \\
\label{Eq:thetadot}\dot{\theta}^2&=&\frac{1}{r^4}\left[{\cal K}-\frac{{\cal L}^2}{\sin^2\theta}\right]
\geq 0\ .
\end{eqnarray}
Based on the equations of motion, one can write the timelike radial motion of massive particles orbiting the BH as follows:
\begin{eqnarray}\label{Eq:rdot}
\dot{r}^2+V_{\rm eff}(r)=\mathcal{E}^2\, ,
\end{eqnarray}
where the radial effective potential $V_{\rm eff}(r)$ is defined by 
\begin{eqnarray} \label{Eq:Veff2}
V_{\rm eff}(r) = \left(1-\frac{2M}{r}-32\pi \rho _{s}r_{s}^{3}\sqrt{\frac{r+r_s}{r_s^2\,r}}  \right) \left( 1+\frac{{\cal L}^2}{r^2} \right)\, ,
\end{eqnarray}
which determines the particle's radial motion at the equatorial plane, i.e., $\theta=\pi/2$. Now, we turn to analyze  $V_{\rm eff}(r)$ for gaining a deeper understanding of timelike geodesics of particles orbiting the Schwarzschild BH surrounded by the Dehnen-type DM halo. 

We shall now analyze the behavior of the effective potential, $V_{\rm eff}$, for the timelike motion of particles orbiting around the Schwarzschild-like BH surrounded by the Dehnen-type DM halo. In Fig.~\ref{fig:eff_pot}, we show the radial profile of $V_{\rm eff}(r)$ for the timelike motion. From the figure, one can observe that the left panel reflects the role of the DM profile parameter $\rho_s$ on the effective potential behavior, while the right panel reflects the effect of the halo core radius parameter $r_s$ for various possible scenarios. It is obvious from the plots of the effective potential that the curves slightly shift towards the right to possibly larger $r$ while its maximum shifts downward with an increasing parameter $\rho_s$, thereby causing the gravitational barrier to weaken.  When considering the effects of parameter $r_s$ ranging from 0.1 to 0.3 with equal intervals, corresponding to the curves from up (blue) to down (purple) for the fixed $\rho_s$, the effective potential changes at a similar rate, resulting in the curves of $V_{\rm eff}$ shifting to the right to possibly larger $r$, as seen in the right panel of Fig.~\ref{fig:eff_pot}. It is important to note that circular orbits that exist around the BH shift towards the right to larger $r$ from the horizon under the combined effects of both DM halo parameters $\rho_s$ and $r_s$. 
\begin{figure*}
\begin{center}
\begin{tabular}{c}
  \includegraphics[scale=0.6]{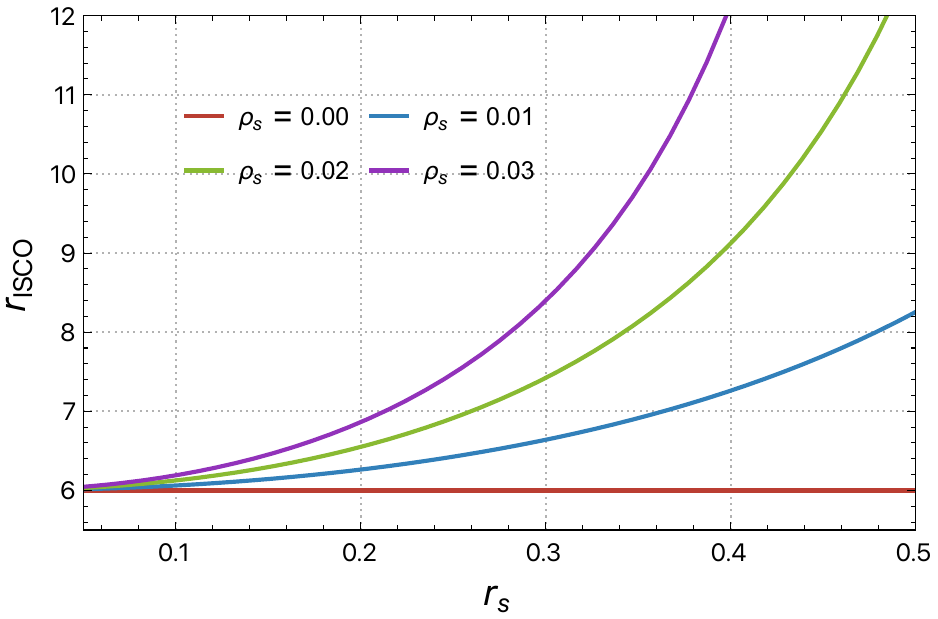}\hspace{0cm}
\end{tabular}
	\caption{\label{fig:isco} 
The ISCO radius of the timelike particles orbiting the
Schwarzschild-like BH surrounded by the Dehnen-type DM halo is plotted against the halo core radius parameter $r_s$ for various values of the DM density profile parameter $\rho_s$. }
\end{center}
\end{figure*}

We then turn to the study circular orbits around the BH surrounded by the DM halo. For the timelike particles to be on the circular orbits, the following required conditions should be satisfied simultaneously  
\begin{eqnarray}\label{Eq:circular}
V_{\rm eff}(r)=\mathcal{E}^2 \mbox{~~and~~}  V_{\rm eff}(r)_{,r}=0\, ,
\end{eqnarray}
where "${,r}$" refers to a derivative with respect to $r$. Eq.~(\ref{Eq:circular}) solves to give the specific angular momentum, $\mathcal{L}$, of timelike particles orbiting on circular orbits, which is given by
\begin{eqnarray}
\mathcal{L}^2&=& \frac{2 r^3 f(r) f(r)_{,r}- r^4f(r)_{,r}^2}{4 r^2 f(r)^2-4 r f(r) f(r)_{,r}+f(r)_{,r}^2}\, .
\end{eqnarray}
However, the angular momentum $\mathcal{L}$ allows the effective potential to have extremum points, leading to timelike particles orbiting on circular orbits. It is worth noting, however, that the coincidence of the maximum and minimum of $V_{eff}$ can only occur at the innermost stable circular orbits (ISCOs). For the timelike particles to be on bound orbits, its specific energy $\mathcal{E}$ and angular momentum $\mathcal{L}$ satisfy \cite{Dadhich22a,Dadhich22IJMPD}  
\begin{eqnarray}\label{range} 
\mathcal{E}_{\rm{ISCO}} \leq \mathcal{E} \leq 1 ~~~~\mbox{and}~~~~\mathcal{L}_{\rm{ISCO}} \leq \mathcal{L},
\end{eqnarray}
where $\mathcal{E}_{\rm{ISCO}}$ and $\mathcal{L}_{\rm{ISCO}}$ respectively refer to the energy and angular momentum of the timelike particle moving on the ISCO around the BH. Note that orbits correspond to captured and escaping ones when particles move on with $\mathcal{E} < \mathcal{E}_{\rm{ISCO}}$ and $\mathcal{E} > 1$, respectively. To gain a deeper understanding of these orbits mentioned and to provide some insights into their behavior, we delve into the properties of the DM halo parameters and their effects on the timelike particle geodesics. Therefore, we first qualitatively analyze the timelike particle geodesics observed from the polar view (i.e., $z = 0$) around the BH and demonstrate their behavior in Fig.~\ref{fig:timelike}. It is worth noting that for simplicity, we shall focus on  the case in which the timelike particles move on orbits restricted to the equatorial plane around the Schwarzschild-like BH surrounded by the Dehnen-type DM halo. As can be seen from Fig.~\ref{fig:timelike}, we observe various orbits, such as escaping, bound and captured orbits, depending on the DM halo parameters. This analysis enhances our understanding of the impacts of these parameters in the vicinity of the BH surrounded by the DM halo. From the timelike particle geodesics, as shown in the top row of Fig.~\ref{fig:timelike}, one can observe that the orbits are initially escaping ones. However, they turn to be bounded and then eventually captured as $\rho_m$ slightly increases while keeping $r_s$ fixed. Similarly, we can observe similar orbits under the effect of the central halo radius $r_s$ for the fixed $\rho_s$. That is, the orbits change from escaping to chaotic behavior and then can slightly be converted into bounded orbits with an increasing parameter $r_s$ ranging from 0.1 to 0.25 with equal intervals, as depicted in the bottom row of Fig.~\ref{fig:timelike}.

To determine the ISCO, one needs to solve the following condition
    \begin{eqnarray}       
    V_{eff(r)_{,rr}}=0\, ,\label{Eq:ddVeff}
    \end{eqnarray}
providing the minimum orbit particles can move on. We now explore the effects of parameters $\rho_s$ and $r_s$ on the ISCO radius where timelike particles can move around the BH. In Fig.~\ref{fig:isco}, we demonstrate the ISCO radii as a function of the parameter $r_s$ for different combinations of the parameter $\rho_s$. It is evident from the figure that the ISCO increases and reaches its maximum possible values with an increasing parameter $r_s$. Similarly, $r_{\rm ISCO}$ changes at a similar rate, causing the ISCO curves to shift upward towards possibly larger values, as shown in Fig.~\ref{fig:isco}. Interestingly, one can infer from the ISCO behavior that it increases to reach larger $r$ under the combined effects of both DM halo parameters $\rho_s$ and $r_s$.

\section{Conclusion}\label{sec5}

From the viewpoint of BH theory, the composition of DM and BH may be an intriguing physical system. In this study, we derive a Schwarzschild-like BH metric in a DM halo with the Dehnen-type density distribution $(1,4,5/2)$ and analyze its properties. We first began to examine the singularity structure of this BH at $r = 0$ by analyzing the spacetime curvature invariants. We found that this BH solution represented by the line elements in Eq.~(\ref{m1}) is unique and that at $r = 0$, there is a physical singularity. Next, we investigated the energy conditions, focusing on analyzing their behaviors in Einstein gravity. We showed that the analytical evaluation of the energy conditions yields intriguing results, leading us to believe that all of the energy conditions are well met. Meanwhile, the pictorial aspect (Fig. \ref{ECs}) confirms this feature for the BHs in the DM halo with a Dehnen-type density profile. 

We also studied the timelike geodesics of particles moving around the Schwarzschild-like BH surrounded by the DM halo with the Dehnen-type density distribution. We showed that the effective potential curves slightly shift towards the right to possibly larger $r$ while its maximum shifts downward as the parameters $\rho_s$ and $r_s$ increase. This behavior results in the weakening of the gravitational barrier, as seen in Fig.~\ref{fig:eff_pot}. We can infer that circular orbits around the BH can be extended to larger $r$ as a consequence of the combined effects of the DM halo parameters $\rho_s$ and $r_s$.  Additionally, we qualitatively analyzed the timelike particle trajectories observed from the polar view (i.e., $z = 0$) to gain insights into their behavior under the effects of the DM halo parameters. This is a crucial tool for understanding the behavior of timelike geodesics in gravity under different possible cases, facilitating an enhanced understanding of the effects of DM halo parameters $\rho_s$ and $r_s$ on the particle geodesics around the BH. We showed various orbits, including escaping, bound, and captured orbits, depending on the DM halo parameters. It was shown that the orbits are initially escaping and become bounded, eventually being captured as $\rho_s$ slightly increases. Similarly, these orbits change at a similar rate from escaping to chaotic behavior and then become bounded orbits under the effect of the central halo radius $r_s$ for the fixed $\rho_s$, as depicted in Fig.~\ref{fig:timelike}. 

Finally, as a consequence of the effects of the DM halo parameters $\rho_s$ and $r_s$, we examined the behavior of the ISCO radii. We showed that the ISCO changes at a similar rate under the impacts of these DM halo parameters, resulting in $r_{\rm ISCO}$ increasing to reach larger $r$ with the rise in $\rho_s$ and $r_s$. 

Given the importance of the DM halo surrounding supermassive BHs at the center of host galaxies, our approach would be of primary astrophysical significance. It does not exclude the BH solution with a DM halo within a Dehnen-type density distribution $(1,4,5/2)$ that could serve as an alternative source of DM halo and play a crucial role in providing insights into the nature of the DM halo.

\bibliography{Ref}

\begin{thebibliography}{59}%
\makeatletter
\providecommand \@ifxundefined [1]{%
 \@ifx{#1\undefined}
}%
\providecommand \@ifnum [1]{%
 \ifnum #1\expandafter \@firstoftwo
 \else \expandafter \@secondoftwo
 \fi
}%
\providecommand \@ifx [1]{%
 \ifx #1\expandafter \@firstoftwo
 \else \expandafter \@secondoftwo
 \fi
}%
\providecommand \natexlab [1]{#1}%
\providecommand \enquote  [1]{``#1''}%
\providecommand \bibnamefont  [1]{#1}%
\providecommand \bibfnamefont [1]{#1}%
\providecommand \citenamefont [1]{#1}%
\providecommand \href@noop [0]{\@secondoftwo}%
\providecommand \href [0]{\begingroup \@sanitize@url \@href}%
\providecommand \@href[1]{\@@startlink{#1}\@@href}%
\providecommand \@@href[1]{\endgroup#1\@@endlink}%
\providecommand \@sanitize@url [0]{\catcode `\\12\catcode `\$12\catcode
  `\&12\catcode `\#12\catcode `\^12\catcode `\_12\catcode `\%12\relax}%
\providecommand \@@startlink[1]{}%
\providecommand \@@endlink[0]{}%
\providecommand \url  [0]{\begingroup\@sanitize@url \@url }%
\providecommand \@url [1]{\endgroup\@href {#1}{\urlprefix }}%
\providecommand \urlprefix  [0]{URL }%
\providecommand \Eprint [0]{\href }%
\providecommand \doibase [0]{http://dx.doi.org/}%
\providecommand \selectlanguage [0]{\@gobble}%
\providecommand \bibinfo  [0]{\@secondoftwo}%
\providecommand \bibfield  [0]{\@secondoftwo}%
\providecommand \translation [1]{[#1]}%
\providecommand \BibitemOpen [0]{}%
\providecommand \bibitemStop [0]{}%
\providecommand \bibitemNoStop [0]{.\EOS\space}%
\providecommand \EOS [0]{\spacefactor3000\relax}%
\providecommand \BibitemShut  [1]{\csname bibitem#1\endcsname}%
\let\auto@bib@innerbib\@empty
\bibitem [{\citenamefont {{Rees}}(1984)}]{Rees84ARAA}%
  \BibitemOpen
  \bibfield  {author} {\bibinfo {author} {\bibfnamefont {M.~J.}\ \bibnamefont
  {{Rees}}},\ }\href {\doibase 10.1146/annurev.aa.22.090184.002351} {\bibfield
  {journal} {\bibinfo  {journal} {Annu. Rev. Astron. Astrophys.}\ }\textbf
  {\bibinfo {volume} {22}},\ \bibinfo {pages} {471} (\bibinfo {year}
  {1984})}\BibitemShut {NoStop}%
\bibitem [{\citenamefont {{Kormendy}}\ and\ \citenamefont
  {{Richstone}}(1995)}]{Kormendy95ARAA}%
  \BibitemOpen
  \bibfield  {author} {\bibinfo {author} {\bibfnamefont {J.}~\bibnamefont
  {{Kormendy}}}\ and\ \bibinfo {author} {\bibfnamefont {D.}~\bibnamefont
  {{Richstone}}},\ }\href {\doibase 10.1146/annurev.aa.33.090195.003053}
  {\bibfield  {journal} {\bibinfo  {journal} {Annu. Rev. Astron. Astrophys.}\
  }\textbf {\bibinfo {volume} {33}},\ \bibinfo {pages} {581} (\bibinfo {year}
  {1995})}\BibitemShut {NoStop}%
\bibitem [{\citenamefont {{Bertone}}\ and\ \citenamefont
  {{Tait}}(2018)}]{Bertone18Nature}%
  \BibitemOpen
  \bibfield  {author} {\bibinfo {author} {\bibfnamefont {G.}~\bibnamefont
  {{Bertone}}}\ and\ \bibinfo {author} {\bibfnamefont {T.~M.~P.}\ \bibnamefont
  {{Tait}}},\ }\href {\doibase 10.1038/s41586-018-0542-z} {\bibfield  {journal}
  {\bibinfo  {journal} {Nature}\ }\textbf {\bibinfo {volume} {562}},\ \bibinfo
  {pages} {51} (\bibinfo {year} {2018})},\ \Eprint
  {http://arxiv.org/abs/1810.01668} {arXiv:1810.01668 [astro-ph.CO]}
  \BibitemShut {NoStop}%
\bibitem [{\citenamefont {{Rubin}}\ and\ \citenamefont
  {{Ford}}(1970)}]{Rubin70ApJ}%
  \BibitemOpen
  \bibfield  {author} {\bibinfo {author} {\bibfnamefont {V.~C.}\ \bibnamefont
  {{Rubin}}}\ and\ \bibinfo {author} {\bibfnamefont {J.}~\bibnamefont {{Ford}},
  \bibfnamefont {W.~Kent}},\ }\href {\doibase 10.1086/150317} {\bibfield
  {journal} {\bibinfo  {journal} {Astrophys. J.}\ }\textbf {\bibinfo {volume}
  {159}},\ \bibinfo {pages} {379} (\bibinfo {year} {1970})}\BibitemShut
  {NoStop}%
\bibitem [{\citenamefont {{Corbelli}}\ and\ \citenamefont
  {{Salucci}}(2000)}]{Corbelli00MNRAS}%
  \BibitemOpen
  \bibfield  {author} {\bibinfo {author} {\bibfnamefont {E.}~\bibnamefont
  {{Corbelli}}}\ and\ \bibinfo {author} {\bibfnamefont {P.}~\bibnamefont
  {{Salucci}}},\ }\href {\doibase 10.1046/j.1365-8711.2000.03075.x} {\bibfield
  {journal} {\bibinfo  {journal} {Mon. Not. Roy. Astron. Soc.}\ }\textbf
  {\bibinfo {volume} {311}},\ \bibinfo {pages} {441} (\bibinfo {year}
  {2000})},\ \Eprint {http://arxiv.org/abs/astro-ph/9909252}
  {arXiv:astro-ph/9909252 [astro-ph]} \BibitemShut {NoStop}%
\bibitem [{\citenamefont {{Komatsu}}\ and\ \citenamefont {et~al. {(WMAP
  Collaboration)}}(2011)}]{Komatsu11ApJS}%
  \BibitemOpen
  \bibfield  {author} {\bibinfo {author} {\bibfnamefont {E.}~\bibnamefont
  {{Komatsu}}}\ and\ \bibinfo {author} {\bibnamefont {et~al. {(WMAP
  Collaboration)}}},\ }\href {\doibase 10.1088/0067-0049/192/2/18} {\bibfield
  {journal} {\bibinfo  {journal} {Astrophys. J. Suppl.}\ }\textbf {\bibinfo
  {volume} {192}},\ \bibinfo {eid} {18} (\bibinfo {year} {2011})},\ \Eprint
  {http://arxiv.org/abs/1001.4538} {arXiv:1001.4538 [astro-ph.CO]} \BibitemShut
  {NoStop}%
\bibitem [{\citenamefont {{Davis}}\ \emph {et~al.}(1985)\citenamefont
  {{Davis}}, \citenamefont {{Efstathiou}}, \citenamefont {{Frenk}},\ and\
  \citenamefont {{White}}}]{Davis85ApJ}%
  \BibitemOpen
  \bibfield  {author} {\bibinfo {author} {\bibfnamefont {M.}~\bibnamefont
  {{Davis}}}, \bibinfo {author} {\bibfnamefont {G.}~\bibnamefont
  {{Efstathiou}}}, \bibinfo {author} {\bibfnamefont {C.~S.}\ \bibnamefont
  {{Frenk}}}, \ and\ \bibinfo {author} {\bibfnamefont {S.~D.~M.}\ \bibnamefont
  {{White}}},\ }\href {\doibase 10.1086/163168} {\bibfield  {journal} {\bibinfo
   {journal} {Astrophys. J.}\ }\textbf {\bibinfo {volume} {292}},\ \bibinfo
  {pages} {371} (\bibinfo {year} {1985})}\BibitemShut {NoStop}%
\bibitem [{\citenamefont {{B{\oe}hm}}\ and\ \citenamefont
  {{Fayet}}(2004)}]{Boehm04NPB}%
  \BibitemOpen
  \bibfield  {author} {\bibinfo {author} {\bibfnamefont {C.}~\bibnamefont
  {{B{\oe}hm}}}\ and\ \bibinfo {author} {\bibfnamefont {P.}~\bibnamefont
  {{Fayet}}},\ }\href {\doibase 10.1016/j.nuclphysb.2004.01.015} {\bibfield
  {journal} {\bibinfo  {journal} {Nucl. Phys. B}\ }\textbf {\bibinfo {volume}
  {683}},\ \bibinfo {pages} {219} (\bibinfo {year} {2004})},\ \Eprint
  {http://arxiv.org/abs/hep-ph/0305261} {arXiv:hep-ph/0305261 [hep-ph]}
  \BibitemShut {NoStop}%
\bibitem [{\citenamefont {{Bertone}}\ \emph
  {et~al.}(2005{\natexlab{a}})\citenamefont {{Bertone}}, \citenamefont
  {{Hooper}},\ and\ \citenamefont {{Silk}}}]{Bertone05PhR}%
  \BibitemOpen
  \bibfield  {author} {\bibinfo {author} {\bibfnamefont {G.}~\bibnamefont
  {{Bertone}}}, \bibinfo {author} {\bibfnamefont {D.}~\bibnamefont {{Hooper}}},
  \ and\ \bibinfo {author} {\bibfnamefont {J.}~\bibnamefont {{Silk}}},\ }\href
  {\doibase 10.1016/j.physrep.2004.08.031} {\bibfield  {journal} {\bibinfo
  {journal} {Phys. Rep.}\ }\textbf {\bibinfo {volume} {405}},\ \bibinfo {pages}
  {279} (\bibinfo {year} {2005}{\natexlab{a}})},\ \Eprint
  {http://arxiv.org/abs/hep-ph/0404175} {arXiv:hep-ph/0404175 [hep-ph]}
  \BibitemShut {NoStop}%
\bibitem [{\citenamefont {{Feng}}\ \emph {et~al.}(2009)\citenamefont {{Feng}},
  \citenamefont {{Kaplinghat}}, \citenamefont {{Tu}},\ and\ \citenamefont
  {{Yu}}}]{Feng09JCAP}%
  \BibitemOpen
  \bibfield  {author} {\bibinfo {author} {\bibfnamefont {J.~L.}\ \bibnamefont
  {{Feng}}}, \bibinfo {author} {\bibfnamefont {M.}~\bibnamefont
  {{Kaplinghat}}}, \bibinfo {author} {\bibfnamefont {H.}~\bibnamefont {{Tu}}},
  \ and\ \bibinfo {author} {\bibfnamefont {H.-B.}\ \bibnamefont {{Yu}}},\
  }\href {\doibase 10.1088/1475-7516/2009/07/004} {\bibfield  {journal}
  {\bibinfo  {journal} {J. Cosmol. Astropart. Phys.}\ }\textbf {\bibinfo
  {volume} {2009}},\ \bibinfo {eid} {004} (\bibinfo {year} {2009})},\ \Eprint
  {http://arxiv.org/abs/0905.3039} {arXiv:0905.3039 [hep-ph]} \BibitemShut
  {NoStop}%
\bibitem [{\citenamefont {{Schumann}}(2019)}]{Schumann19}%
  \BibitemOpen
  \bibfield  {author} {\bibinfo {author} {\bibfnamefont {M.}~\bibnamefont
  {{Schumann}}},\ }\href {\doibase 10.1088/1361-6471/ab2ea5} {\bibfield
  {journal} {\bibinfo  {journal} {J. Phys. G Nucl. Part. Phys.}\ }\textbf
  {\bibinfo {volume} {46}},\ \bibinfo {pages} {103003} (\bibinfo {year}
  {2019})},\ \Eprint {http://arxiv.org/abs/1903.03026} {arXiv:1903.03026
  [astro-ph.CO]} \BibitemShut {NoStop}%
\bibitem [{\citenamefont {Sofue}(2020)}]{Sofue2020RotationCO}%
  \BibitemOpen
  \bibfield  {author} {\bibinfo {author} {\bibfnamefont {Y.}~\bibnamefont
  {Sofue}},\ }\href {https://api.semanticscholar.org/CorpusID:216144423}
  {\bibfield  {journal} {\bibinfo  {journal} {Galaxies}\ } (\bibinfo {year}
  {2020})}\BibitemShut {NoStop}%
\bibitem [{\citenamefont {{Clowe}}\ \emph {et~al.}(2006)\citenamefont
  {{Clowe}}, \citenamefont {{Brada{\v{c}}}}, \citenamefont {{Gonzalez}},
  \citenamefont {{Markevitch}}, \citenamefont {{Randall}}, \citenamefont
  {{Jones}},\ and\ \citenamefont {{Zaritsky}}}]{Clowe06ApJL}%
  \BibitemOpen
  \bibfield  {author} {\bibinfo {author} {\bibfnamefont {D.}~\bibnamefont
  {{Clowe}}}, \bibinfo {author} {\bibfnamefont {M.}~\bibnamefont
  {{Brada{\v{c}}}}}, \bibinfo {author} {\bibfnamefont {A.~H.}\ \bibnamefont
  {{Gonzalez}}}, \bibinfo {author} {\bibfnamefont {M.}~\bibnamefont
  {{Markevitch}}}, \bibinfo {author} {\bibfnamefont {S.~W.}\ \bibnamefont
  {{Randall}}}, \bibinfo {author} {\bibfnamefont {C.}~\bibnamefont {{Jones}}},
  \ and\ \bibinfo {author} {\bibfnamefont {D.}~\bibnamefont {{Zaritsky}}},\
  }\href {\doibase 10.1086/508162} {\bibfield  {journal} {\bibinfo  {journal}
  {Astrophys. J. Lett.}\ }\textbf {\bibinfo {volume} {648}},\ \bibinfo {pages}
  {L109} (\bibinfo {year} {2006})},\ \Eprint
  {http://arxiv.org/abs/astro-ph/0608407} {arXiv:astro-ph/0608407 [astro-ph]}
  \BibitemShut {NoStop}%
\bibitem [{\citenamefont {{Jusufi}}\ \emph {et~al.}(2020)\citenamefont
  {{Jusufi}}, \citenamefont {{Jamil}},\ and\ \citenamefont
  {{Zhu}}}]{Jusufi20EPJC}%
  \BibitemOpen
  \bibfield  {author} {\bibinfo {author} {\bibfnamefont {K.}~\bibnamefont
  {{Jusufi}}}, \bibinfo {author} {\bibfnamefont {M.}~\bibnamefont {{Jamil}}}, \
  and\ \bibinfo {author} {\bibfnamefont {T.}~\bibnamefont {{Zhu}}},\ }\href
  {\doibase 10.1140/epjc/s10052-020-7899-5} {\bibfield  {journal} {\bibinfo
  {journal} {Eur. Phys. J. C}\ }\textbf {\bibinfo {volume} {80}},\ \bibinfo
  {eid} {354} (\bibinfo {year} {2020})},\ \Eprint
  {http://arxiv.org/abs/2005.05299} {arXiv:2005.05299 [gr-qc]} \BibitemShut
  {NoStop}%
\bibitem [{\citenamefont {Das}\ \emph {et~al.}(2022)\citenamefont {Das},
  \citenamefont {Saha},\ and\ \citenamefont {Gangopadhyay}}]{Das_2022}%
  \BibitemOpen
  \bibfield  {author} {\bibinfo {author} {\bibfnamefont {A.}~\bibnamefont
  {Das}}, \bibinfo {author} {\bibfnamefont {A.}~\bibnamefont {Saha}}, \ and\
  \bibinfo {author} {\bibfnamefont {S.}~\bibnamefont {Gangopadhyay}},\ }\href
  {\doibase 10.1088/1361-6382/ac50ed} {\bibfield  {journal} {\bibinfo
  {journal} {Classical and Quantum Gravity}\ }\textbf {\bibinfo {volume}
  {39}},\ \bibinfo {pages} {075005} (\bibinfo {year} {2022})}\BibitemShut
  {NoStop}%
\bibitem [{\citenamefont {{Karamazov}}\ \emph {et~al.}(2021)\citenamefont
  {{Karamazov}}, \citenamefont {{Timko}},\ and\ \citenamefont
  {{Heyrovsk{\'y}}}}]{Karamazov21ApJ}%
  \BibitemOpen
  \bibfield  {author} {\bibinfo {author} {\bibfnamefont {M.}~\bibnamefont
  {{Karamazov}}}, \bibinfo {author} {\bibfnamefont {L.}~\bibnamefont
  {{Timko}}}, \ and\ \bibinfo {author} {\bibfnamefont {D.}~\bibnamefont
  {{Heyrovsk{\'y}}}},\ }\href {\doibase 10.3847/1538-4357/ac151c} {\bibfield
  {journal} {\bibinfo  {journal} {Astrophys. J.}\ }\textbf {\bibinfo {volume}
  {922}},\ \bibinfo {eid} {72} (\bibinfo {year} {2021})},\ \Eprint
  {http://arxiv.org/abs/2103.16965} {arXiv:2103.16965 [astro-ph.GA]}
  \BibitemShut {NoStop}%
\bibitem [{\citenamefont {{Qi, Qi}}\ \emph {et~al.}(2023)\citenamefont {{Qi,
  Qi}}, \citenamefont {{Meng, Yuan}}, \citenamefont {{Wang, Xi-Jing}},\ and\
  \citenamefont {{Kuang, Xiao-Mei}}}]{Qi23}%
  \BibitemOpen
  \bibfield  {author} {\bibinfo {author} {\bibnamefont {{Qi, Qi}}}, \bibinfo
  {author} {\bibnamefont {{Meng, Yuan}}}, \bibinfo {author} {\bibnamefont
  {{Wang, Xi-Jing}}}, \ and\ \bibinfo {author} {\bibnamefont {{Kuang,
  Xiao-Mei}}},\ }\href {\doibase 10.1140/epjc/s10052-023-12233-z} {\bibfield
  {journal} {\bibinfo  {journal} {Eur. Phys. J. C}\ }\textbf {\bibinfo {volume}
  {83}},\ \bibinfo {pages} {1043} (\bibinfo {year} {2023})}\BibitemShut
  {NoStop}%
\bibitem [{\citenamefont {{Bertone}}\ \emph
  {et~al.}(2005{\natexlab{b}})\citenamefont {{Bertone}}, \citenamefont
  {{Hooper}},\ and\ \citenamefont {{Silk}}}]{Bertone05}%
  \BibitemOpen
  \bibfield  {author} {\bibinfo {author} {\bibfnamefont {G.}~\bibnamefont
  {{Bertone}}}, \bibinfo {author} {\bibfnamefont {D.}~\bibnamefont {{Hooper}}},
  \ and\ \bibinfo {author} {\bibfnamefont {J.}~\bibnamefont {{Silk}}},\ }\href
  {\doibase 10.1016/j.physrep.2004.08.031} {\bibfield  {journal} {\bibinfo
  {journal} {Phys. Rep.}\ }\textbf {\bibinfo {volume} {405}},\ \bibinfo {pages}
  {279} (\bibinfo {year} {2005}{\natexlab{b}})},\ \Eprint
  {http://arxiv.org/abs/hep-ph/0404175} {arXiv:hep-ph/0404175 [hep-ph]}
  \BibitemShut {NoStop}%
\bibitem [{\citenamefont {{de Swart}}\ \emph {et~al.}(2017)\citenamefont {{de
  Swart}}, \citenamefont {{Bertone}},\ and\ \citenamefont {{van
  Dongen}}}]{deSwart17Nat}%
  \BibitemOpen
  \bibfield  {author} {\bibinfo {author} {\bibfnamefont {J.~G.}\ \bibnamefont
  {{de Swart}}}, \bibinfo {author} {\bibfnamefont {G.}~\bibnamefont
  {{Bertone}}}, \ and\ \bibinfo {author} {\bibfnamefont {J.}~\bibnamefont {{van
  Dongen}}},\ }\href {\doibase 10.1038/s41550-017-0059} {\bibfield  {journal}
  {\bibinfo  {journal} {Nature Astron.}\ }\textbf {\bibinfo {volume} {1}},\
  \bibinfo {eid} {0059} (\bibinfo {year} {2017})},\ \Eprint
  {http://arxiv.org/abs/1703.00013} {arXiv:1703.00013 [astro-ph.CO]}
  \BibitemShut {NoStop}%
\bibitem [{\citenamefont {{Wechsler}}\ and\ \citenamefont
  {{Tinker}}(2018)}]{Wechsler18}%
  \BibitemOpen
  \bibfield  {author} {\bibinfo {author} {\bibfnamefont {R.~H.}\ \bibnamefont
  {{Wechsler}}}\ and\ \bibinfo {author} {\bibfnamefont {J.~L.}\ \bibnamefont
  {{Tinker}}},\ }\href {\doibase 10.1146/annurev-astro-081817-051756}
  {\bibfield  {journal} {\bibinfo  {journal} {Annu. Rev. Astron. Astrophys.}\
  }\textbf {\bibinfo {volume} {56}},\ \bibinfo {pages} {435} (\bibinfo {year}
  {2018})},\ \Eprint {http://arxiv.org/abs/1804.03097} {arXiv:1804.03097
  [astro-ph.GA]} \BibitemShut {NoStop}%
\bibitem [{\citenamefont {{Valluri}}\ \emph {et~al.}(2004)\citenamefont
  {{Valluri}}, \citenamefont {{Merritt}},\ and\ \citenamefont
  {{Emsellem}}}]{Valluri04ApJ}%
  \BibitemOpen
  \bibfield  {author} {\bibinfo {author} {\bibfnamefont {M.}~\bibnamefont
  {{Valluri}}}, \bibinfo {author} {\bibfnamefont {D.}~\bibnamefont
  {{Merritt}}}, \ and\ \bibinfo {author} {\bibfnamefont {E.}~\bibnamefont
  {{Emsellem}}},\ }\href {\doibase 10.1086/380896} {\bibfield  {journal}
  {\bibinfo  {journal} {Astrophys. J.}\ }\textbf {\bibinfo {volume} {602}},\
  \bibinfo {pages} {66} (\bibinfo {year} {2004})},\ \Eprint
  {http://arxiv.org/abs/astro-ph/0210379} {arXiv:astro-ph/0210379 [astro-ph]}
  \BibitemShut {NoStop}%
\bibitem [{\citenamefont {{Akiyama}}\ and\ \citenamefont {et~al. {(Event
  Horizon Telescope Collaboration)}}(2019{\natexlab{a}})}]{Akiyama19L1}%
  \BibitemOpen
  \bibfield  {author} {\bibinfo {author} {\bibfnamefont {K.}~\bibnamefont
  {{Akiyama}}}\ and\ \bibinfo {author} {\bibnamefont {et~al. {(Event Horizon
  Telescope Collaboration)}}},\ }\href {\doibase 10.3847/2041-8213/ab0ec7}
  {\bibfield  {journal} {\bibinfo  {journal} {Astrophys. J.}\ }\textbf
  {\bibinfo {volume} {875}},\ \bibinfo {eid} {L1} (\bibinfo {year}
  {2019}{\natexlab{a}})},\ \Eprint {http://arxiv.org/abs/1906.11238}
  {arXiv:1906.11238 [astro-ph.GA]} \BibitemShut {NoStop}%
\bibitem [{\citenamefont {{Akiyama}}\ and\ \citenamefont {et~al. {(Event
  Horizon Telescope Collaboration)}}(2019{\natexlab{b}})}]{Akiyama19L6}%
  \BibitemOpen
  \bibfield  {author} {\bibinfo {author} {\bibfnamefont {K.}~\bibnamefont
  {{Akiyama}}}\ and\ \bibinfo {author} {\bibnamefont {et~al. {(Event Horizon
  Telescope Collaboration)}}},\ }\href {\doibase 10.3847/2041-8213/ab1141}
  {\bibfield  {journal} {\bibinfo  {journal} {Astrophys. J.}\ }\textbf
  {\bibinfo {volume} {875}},\ \bibinfo {eid} {L6} (\bibinfo {year}
  {2019}{\natexlab{b}})},\ \Eprint {http://arxiv.org/abs/1906.11243}
  {arXiv:1906.11243 [astro-ph.GA]} \BibitemShut {NoStop}%
\bibitem [{\citenamefont {{Akiyama}}\ and\ \citenamefont {et~al. {(Event
  Horizon Telescope Collaboration)}}(2022)}]{Akiyama22L12}%
  \BibitemOpen
  \bibfield  {author} {\bibinfo {author} {\bibfnamefont {K.}~\bibnamefont
  {{Akiyama}}}\ and\ \bibinfo {author} {\bibnamefont {et~al. {(Event Horizon
  Telescope Collaboration)}}},\ }\href {\doibase 10.3847/2041-8213/ac6674}
  {\bibfield  {journal} {\bibinfo  {journal} {Astrophys. J. Lett.}\ }\textbf
  {\bibinfo {volume} {930}},\ \bibinfo {eid} {L12} (\bibinfo {year}
  {2022})}\BibitemShut {NoStop}%
\bibitem [{\citenamefont {{Persic}}\ \emph {et~al.}(1996)\citenamefont
  {{Persic}}, \citenamefont {{Salucci}},\ and\ \citenamefont
  {{Stel}}}]{Persic96}%
  \BibitemOpen
  \bibfield  {author} {\bibinfo {author} {\bibfnamefont {M.}~\bibnamefont
  {{Persic}}}, \bibinfo {author} {\bibfnamefont {P.}~\bibnamefont {{Salucci}}},
  \ and\ \bibinfo {author} {\bibfnamefont {F.}~\bibnamefont {{Stel}}},\ }\href
  {\doibase 10.1093/mnras/278.1.27} {\bibfield  {journal} {\bibinfo  {journal}
  {Mon. Not. R. Astron. Soc.}\ }\textbf {\bibinfo {volume} {281}},\ \bibinfo
  {pages} {27} (\bibinfo {year} {1996})},\ \Eprint
  {http://arxiv.org/abs/astro-ph/9506004} {arXiv:astro-ph/9506004 [astro-ph]}
  \BibitemShut {NoStop}%
\bibitem [{\citenamefont {{Li}}\ and\ \citenamefont
  {{Yang}}(2012)}]{Li-Yang12}%
  \BibitemOpen
  \bibfield  {author} {\bibinfo {author} {\bibfnamefont {M.-H.}\ \bibnamefont
  {{Li}}}\ and\ \bibinfo {author} {\bibfnamefont {K.-C.}\ \bibnamefont
  {{Yang}}},\ }\href {\doibase 10.1103/PhysRevD.86.123015} {\bibfield
  {journal} {\bibinfo  {journal} {Phys. Rev. D}\ }\textbf {\bibinfo {volume}
  {86}},\ \bibinfo {eid} {123015} (\bibinfo {year} {2012})},\ \Eprint
  {http://arxiv.org/abs/1204.3178} {arXiv:1204.3178 [astro-ph.CO]} \BibitemShut
  {NoStop}%
\bibitem [{\citenamefont {{Hendi}}\ \emph {et~al.}(2020)\citenamefont
  {{Hendi}}, \citenamefont {{Nemati}}, \citenamefont {{Lin}},\ and\
  \citenamefont {{Jamil}}}]{Hendi20}%
  \BibitemOpen
  \bibfield  {author} {\bibinfo {author} {\bibfnamefont {S.~H.}\ \bibnamefont
  {{Hendi}}}, \bibinfo {author} {\bibfnamefont {A.}~\bibnamefont {{Nemati}}},
  \bibinfo {author} {\bibfnamefont {K.}~\bibnamefont {{Lin}}}, \ and\ \bibinfo
  {author} {\bibfnamefont {M.}~\bibnamefont {{Jamil}}},\ }\href {\doibase
  10.1140/epjc/s10052-020-7829-6} {\bibfield  {journal} {\bibinfo  {journal}
  {Eur. Phys. J. C}\ }\textbf {\bibinfo {volume} {80}},\ \bibinfo {eid} {296}
  (\bibinfo {year} {2020})},\ \Eprint {http://arxiv.org/abs/2001.01591}
  {arXiv:2001.01591 [gr-qc]} \BibitemShut {NoStop}%
\bibitem [{\citenamefont {{Rizwan}}\ \emph {et~al.}(2019)\citenamefont
  {{Rizwan}}, \citenamefont {{Jamil}},\ and\ \citenamefont
  {{Jusufi}}}]{Rizwan19}%
  \BibitemOpen
  \bibfield  {author} {\bibinfo {author} {\bibfnamefont {M.}~\bibnamefont
  {{Rizwan}}}, \bibinfo {author} {\bibfnamefont {M.}~\bibnamefont {{Jamil}}}, \
  and\ \bibinfo {author} {\bibfnamefont {K.}~\bibnamefont {{Jusufi}}},\ }\href
  {\doibase 10.1103/PhysRevD.99.024050} {\bibfield  {journal} {\bibinfo
  {journal} {Phys. Rev. D}\ }\textbf {\bibinfo {volume} {99}},\ \bibinfo {eid}
  {024050} (\bibinfo {year} {2019})},\ \Eprint
  {http://arxiv.org/abs/1812.01331} {arXiv:1812.01331 [gr-qc]} \BibitemShut
  {NoStop}%
\bibitem [{\citenamefont {{Shaymatov}}\ \emph
  {et~al.}(2021{\natexlab{a}})\citenamefont {{Shaymatov}}, \citenamefont
  {{Ahmedov}},\ and\ \citenamefont {{Jamil}}}]{Shaymatov21d}%
  \BibitemOpen
  \bibfield  {author} {\bibinfo {author} {\bibfnamefont {S.}~\bibnamefont
  {{Shaymatov}}}, \bibinfo {author} {\bibfnamefont {B.}~\bibnamefont
  {{Ahmedov}}}, \ and\ \bibinfo {author} {\bibfnamefont {M.}~\bibnamefont
  {{Jamil}}},\ }\href {\doibase 10.1140/epjc/s10052-021-09398-w} {\bibfield
  {journal} {\bibinfo  {journal} {Eur. Phys. J. C}\ }\textbf {\bibinfo {volume}
  {81}},\ \bibinfo {eid} {588} (\bibinfo {year}
  {2021}{\natexlab{a}})}\BibitemShut {NoStop}%
\bibitem [{\citenamefont {{Rayimbaev}}\ \emph {et~al.}(2021)\citenamefont
  {{Rayimbaev}}, \citenamefont {{Shaymatov}},\ and\ \citenamefont
  {{Jamil}}}]{Rayimbaev-Shaymatov21a}%
  \BibitemOpen
  \bibfield  {author} {\bibinfo {author} {\bibfnamefont {J.}~\bibnamefont
  {{Rayimbaev}}}, \bibinfo {author} {\bibfnamefont {S.}~\bibnamefont
  {{Shaymatov}}}, \ and\ \bibinfo {author} {\bibfnamefont {M.}~\bibnamefont
  {{Jamil}}},\ }\href {\doibase 10.1140/epjc/s10052-021-09488-9} {\bibfield
  {journal} {\bibinfo  {journal} {Eur. Phys. J. C}\ }\textbf {\bibinfo {volume}
  {81}},\ \bibinfo {eid} {699} (\bibinfo {year} {2021})},\ \Eprint
  {http://arxiv.org/abs/2107.13436} {arXiv:2107.13436 [gr-qc]} \BibitemShut
  {NoStop}%
\bibitem [{\citenamefont {{Shaymatov}}\ \emph
  {et~al.}(2021{\natexlab{b}})\citenamefont {{Shaymatov}}, \citenamefont
  {{Malafarina}},\ and\ \citenamefont {{Ahmedov}}}]{Shaymatov21pdu}%
  \BibitemOpen
  \bibfield  {author} {\bibinfo {author} {\bibfnamefont {S.}~\bibnamefont
  {{Shaymatov}}}, \bibinfo {author} {\bibfnamefont {D.}~\bibnamefont
  {{Malafarina}}}, \ and\ \bibinfo {author} {\bibfnamefont {B.}~\bibnamefont
  {{Ahmedov}}},\ }\href {\doibase 10.1016/j.dark.2021.100891} {\bibfield
  {journal} {\bibinfo  {journal} {Phys. Dark Universe}\ }\textbf {\bibinfo
  {volume} {34}},\ \bibinfo {eid} {100891} (\bibinfo {year}
  {2021}{\natexlab{b}})},\ \Eprint {http://arxiv.org/abs/2004.06811}
  {arXiv:2004.06811 [gr-qc]} \BibitemShut {NoStop}%
\bibitem [{\citenamefont {{Shaymatov}}\ \emph {et~al.}(2022)\citenamefont
  {{Shaymatov}}, \citenamefont {{Sheoran}},\ and\ \citenamefont
  {{Siwach}}}]{Shaymatov22a}%
  \BibitemOpen
  \bibfield  {author} {\bibinfo {author} {\bibfnamefont {S.}~\bibnamefont
  {{Shaymatov}}}, \bibinfo {author} {\bibfnamefont {P.}~\bibnamefont
  {{Sheoran}}}, \ and\ \bibinfo {author} {\bibfnamefont {S.}~\bibnamefont
  {{Siwach}}},\ }\href {\doibase 10.1103/PhysRevD.105.104059} {\bibfield
  {journal} {\bibinfo  {journal} {Phys. Rev. D}\ }\textbf {\bibinfo {volume}
  {105}},\ \bibinfo {eid} {104059} (\bibinfo {year} {2022})},\ \Eprint
  {http://arxiv.org/abs/2110.10610} {arXiv:2110.10610 [gr-qc]} \BibitemShut
  {NoStop}%
\bibitem [{\citenamefont {{Cardoso}}\ \emph {et~al.}(2022)\citenamefont
  {{Cardoso}}, \citenamefont {{Destounis}}, \citenamefont {{Duque}},
  \citenamefont {{Macedo}},\ and\ \citenamefont {{Maselli}}}]{Cardoso22DM}%
  \BibitemOpen
  \bibfield  {author} {\bibinfo {author} {\bibfnamefont {V.}~\bibnamefont
  {{Cardoso}}}, \bibinfo {author} {\bibfnamefont {K.}~\bibnamefont
  {{Destounis}}}, \bibinfo {author} {\bibfnamefont {F.}~\bibnamefont
  {{Duque}}}, \bibinfo {author} {\bibfnamefont {R.~P.}\ \bibnamefont
  {{Macedo}}}, \ and\ \bibinfo {author} {\bibfnamefont {A.}~\bibnamefont
  {{Maselli}}},\ }\href {\doibase 10.1103/PhysRevD.105.L061501} {\bibfield
  {journal} {\bibinfo  {journal} {Phys. Rev. D}\ }\textbf {\bibinfo {volume}
  {105}},\ \bibinfo {eid} {L061501} (\bibinfo {year} {2022})},\ \Eprint
  {http://arxiv.org/abs/2109.00005} {arXiv:2109.00005 [gr-qc]} \BibitemShut
  {NoStop}%
\bibitem [{\citenamefont {{Hou}}\ \emph
  {et~al.}(2018{\natexlab{a}})\citenamefont {{Hou}}, \citenamefont {{Xu}},
  \citenamefont {{Zhou}},\ and\ \citenamefont {{Wang}}}]{Hou18-dm}%
  \BibitemOpen
  \bibfield  {author} {\bibinfo {author} {\bibfnamefont {X.}~\bibnamefont
  {{Hou}}}, \bibinfo {author} {\bibfnamefont {Z.}~\bibnamefont {{Xu}}},
  \bibinfo {author} {\bibfnamefont {M.}~\bibnamefont {{Zhou}}}, \ and\ \bibinfo
  {author} {\bibfnamefont {J.}~\bibnamefont {{Wang}}},\ }\href {\doibase
  10.1088/1475-7516/2018/07/015} {\bibfield  {journal} {\bibinfo  {journal} {J.
  Cosmol. Astropart. Phys.}\ }\textbf {\bibinfo {volume} {2018}},\ \bibinfo
  {eid} {015} (\bibinfo {year} {2018}{\natexlab{a}})},\ \Eprint
  {http://arxiv.org/abs/1804.08110} {arXiv:1804.08110 [gr-qc]} \BibitemShut
  {NoStop}%
\bibitem [{\citenamefont {{Shen}}\ \emph {et~al.}(2024)\citenamefont {{Shen}},
  \citenamefont {{Wang}}, \citenamefont {{Gong}},\ and\ \citenamefont
  {{Yin}}}]{Shen24PLB}%
  \BibitemOpen
  \bibfield  {author} {\bibinfo {author} {\bibfnamefont {Z.}~\bibnamefont
  {{Shen}}}, \bibinfo {author} {\bibfnamefont {A.}~\bibnamefont {{Wang}}},
  \bibinfo {author} {\bibfnamefont {Y.}~\bibnamefont {{Gong}}}, \ and\ \bibinfo
  {author} {\bibfnamefont {S.}~\bibnamefont {{Yin}}},\ }\href {\doibase
  10.1016/j.physletb.2024.138797} {\bibfield  {journal} {\bibinfo  {journal}
  {Phys. Lett. B}\ }\textbf {\bibinfo {volume} {855}},\ \bibinfo {eid} {138797}
  (\bibinfo {year} {2024})},\ \Eprint {http://arxiv.org/abs/2311.12259}
  {arXiv:2311.12259 [gr-qc]} \BibitemShut {NoStop}%
\bibitem [{\citenamefont {{Navarro}}\ \emph {et~al.}(1996)\citenamefont
  {{Navarro}}, \citenamefont {{Frenk}},\ and\ \citenamefont
  {{White}}}]{Navarro96ApJ}%
  \BibitemOpen
  \bibfield  {author} {\bibinfo {author} {\bibfnamefont {J.~F.}\ \bibnamefont
  {{Navarro}}}, \bibinfo {author} {\bibfnamefont {C.~S.}\ \bibnamefont
  {{Frenk}}}, \ and\ \bibinfo {author} {\bibfnamefont {S.~D.~M.}\ \bibnamefont
  {{White}}},\ }\href {\doibase 10.1086/177173} {\bibfield  {journal} {\bibinfo
   {journal} {Astrophys. J.}\ }\textbf {\bibinfo {volume} {462}},\ \bibinfo
  {pages} {563} (\bibinfo {year} {1996})},\ \Eprint
  {http://arxiv.org/abs/astro-ph/9508025} {arXiv:astro-ph/9508025 [astro-ph]}
  \BibitemShut {NoStop}%
\bibitem [{\citenamefont {Dutton}\ and\ \citenamefont
  {Macci{\`o}}(2014)}]{Dutton14MNRAS}%
  \BibitemOpen
  \bibfield  {author} {\bibinfo {author} {\bibfnamefont {A.}~\bibnamefont
  {Dutton}}\ and\ \bibinfo {author} {\bibfnamefont {A.}~\bibnamefont
  {Macci{\`o}}},\ }\href {\doibase 10.1093/mnras/stu742} {\bibfield  {journal}
  {\bibinfo  {journal} {Mon. Not. Roy. Astron. Soc.}\ }\textbf {\bibinfo
  {volume} {441}},\ \bibinfo {pages} {3359} (\bibinfo {year}
  {2014})}\BibitemShut {NoStop}%
\bibitem [{\citenamefont {{Merritt}}\ \emph {et~al.}(2006)\citenamefont
  {{Merritt}}, \citenamefont {{Graham}}, \citenamefont {{Moore}}, \citenamefont
  {{Diemand}},\ and\ \citenamefont {{Terzi{\'c}}}}]{Merritt06ApJ}%
  \BibitemOpen
  \bibfield  {author} {\bibinfo {author} {\bibfnamefont {D.}~\bibnamefont
  {{Merritt}}}, \bibinfo {author} {\bibfnamefont {A.~W.}\ \bibnamefont
  {{Graham}}}, \bibinfo {author} {\bibfnamefont {B.}~\bibnamefont {{Moore}}},
  \bibinfo {author} {\bibfnamefont {J.}~\bibnamefont {{Diemand}}}, \ and\
  \bibinfo {author} {\bibfnamefont {B.}~\bibnamefont {{Terzi{\'c}}}},\ }\href
  {\doibase 10.1086/508988} {\bibfield  {journal} {\bibinfo  {journal} {Astron.
  J.}\ }\textbf {\bibinfo {volume} {132}},\ \bibinfo {pages} {2685} (\bibinfo
  {year} {2006})},\ \Eprint {http://arxiv.org/abs/astro-ph/0509417}
  {arXiv:astro-ph/0509417 [astro-ph]} \BibitemShut {NoStop}%
\bibitem [{\citenamefont {{Burkert}}(1995)}]{Burkert95ApJL}%
  \BibitemOpen
  \bibfield  {author} {\bibinfo {author} {\bibfnamefont {A.}~\bibnamefont
  {{Burkert}}},\ }\href {\doibase 10.1086/309560} {\bibfield  {journal}
  {\bibinfo  {journal} {Astrophys. J. Lett.}\ }\textbf {\bibinfo {volume}
  {447}},\ \bibinfo {pages} {L25} (\bibinfo {year} {1995})},\ \Eprint
  {http://arxiv.org/abs/astro-ph/9504041} {arXiv:astro-ph/9504041 [astro-ph]}
  \BibitemShut {NoStop}%
\bibitem [{\citenamefont {{Dehnen}}(1993)}]{Dehnen93}%
  \BibitemOpen
  \bibfield  {author} {\bibinfo {author} {\bibfnamefont {W.}~\bibnamefont
  {{Dehnen}}},\ }\href {\doibase 10.1093/mnras/265.1.250} {\bibfield  {journal}
  {\bibinfo  {journal} {Mon. Not. R. Astron. Soc.}\ }\textbf {\bibinfo {volume}
  {265}},\ \bibinfo {pages} {250} (\bibinfo {year} {1993})}\BibitemShut
  {NoStop}%
\bibitem [{\citenamefont {{Shukirgaliyev, B.}}\ \emph
  {et~al.}(2021)\citenamefont {{Shukirgaliyev, B.}}, \citenamefont {{Otebay,
  A.}}, \citenamefont {{Sobolenko, M.}}, \citenamefont {{Ishchenko, M.}},
  \citenamefont {{Borodina, O.}}, \citenamefont {{Panamarev, T.}},
  \citenamefont {{Myrzakul, S.}}, \citenamefont {{Kalambay, M.}}, \citenamefont
  {{Naurzbayeva, A.}}, \citenamefont {{Abdikamalov, E.}}, \citenamefont
  {{Polyachenko, E.}}, \citenamefont {{Banerjee, S.}}, \citenamefont {{Berczik,
  P.}}, \citenamefont {{Spurzem, R.}},\ and\ \citenamefont {{Just,
  A.}}}]{Shukirgaliyev21A&A}%
  \BibitemOpen
  \bibfield  {author} {\bibinfo {author} {\bibnamefont {{Shukirgaliyev, B.}}},
  \bibinfo {author} {\bibnamefont {{Otebay, A.}}}, \bibinfo {author}
  {\bibnamefont {{Sobolenko, M.}}}, \bibinfo {author} {\bibnamefont
  {{Ishchenko, M.}}}, \bibinfo {author} {\bibnamefont {{Borodina, O.}}},
  \bibinfo {author} {\bibnamefont {{Panamarev, T.}}}, \bibinfo {author}
  {\bibnamefont {{Myrzakul, S.}}}, \bibinfo {author} {\bibnamefont {{Kalambay,
  M.}}}, \bibinfo {author} {\bibnamefont {{Naurzbayeva, A.}}}, \bibinfo
  {author} {\bibnamefont {{Abdikamalov, E.}}}, \bibinfo {author} {\bibnamefont
  {{Polyachenko, E.}}}, \bibinfo {author} {\bibnamefont {{Banerjee, S.}}},
  \bibinfo {author} {\bibnamefont {{Berczik, P.}}}, \bibinfo {author}
  {\bibnamefont {{Spurzem, R.}}}, \ and\ \bibinfo {author} {\bibnamefont
  {{Just, A.}}},\ }\href {\doibase 10.1051/0004-6361/202141299} {\bibfield
  {journal} {\bibinfo  {journal} {Astron. Astrophys.}\ }\textbf {\bibinfo
  {volume} {654}},\ \bibinfo {pages} {A53} (\bibinfo {year}
  {2021})}\BibitemShut {NoStop}%
\bibitem [{\citenamefont {{Pantig}}\ and\ \citenamefont
  {{{\"O}vg{\"u}n}}(2022)}]{Pantig22JCAP}%
  \BibitemOpen
  \bibfield  {author} {\bibinfo {author} {\bibfnamefont {R.~C.}\ \bibnamefont
  {{Pantig}}}\ and\ \bibinfo {author} {\bibfnamefont {A.}~\bibnamefont
  {{{\"O}vg{\"u}n}}},\ }\href {\doibase 10.1088/1475-7516/2022/08/056}
  {\bibfield  {journal} {\bibinfo  {journal} {J. Cosmol. Astropart. Phys.}\
  }\textbf {\bibinfo {volume} {2022}},\ \bibinfo {eid} {056} (\bibinfo {year}
  {2022})},\ \Eprint {http://arxiv.org/abs/2202.07404} {arXiv:2202.07404
  [astro-ph.GA]} \BibitemShut {NoStop}%
\bibitem [{\citenamefont {Gohain}\ \emph {et~al.}(2024)\citenamefont {Gohain},
  \citenamefont {Phukon},\ and\ \citenamefont {Bhuyan}}]{Gohain24DM}%
  \BibitemOpen
  \bibfield  {author} {\bibinfo {author} {\bibfnamefont {M.~M.}\ \bibnamefont
  {Gohain}}, \bibinfo {author} {\bibfnamefont {P.}~\bibnamefont {Phukon}}, \
  and\ \bibinfo {author} {\bibfnamefont {K.}~\bibnamefont {Bhuyan}},\ }\href
  {\doibase 10.1016/j.dark.2024.101683} {\bibfield  {journal} {\bibinfo
  {journal} {Phys. Dark Univ.}\ }\textbf {\bibinfo {volume} {46}},\ \bibinfo
  {pages} {101683} (\bibinfo {year} {2024})},\ \Eprint
  {http://arxiv.org/abs/2407.02872} {arXiv:2407.02872 [gr-qc]} \BibitemShut
  {NoStop}%
\bibitem [{\citenamefont {{Matos}}\ and\ \citenamefont
  {{Nunez}}(2005)}]{Matos05}%
  \BibitemOpen
  \bibfield  {author} {\bibinfo {author} {\bibfnamefont {T.}~\bibnamefont
  {{Matos}}}\ and\ \bibinfo {author} {\bibfnamefont {D.}~\bibnamefont
  {{Nunez}}},\ }\href@noop {} {\bibfield  {journal} {\bibinfo  {journal} {Rev.
  Mex. Fis.}\ }\textbf {\bibinfo {volume} {51}},\ \bibinfo {pages} {71}
  (\bibinfo {year} {2005})}\BibitemShut {NoStop}%
\bibitem [{\citenamefont {{Xu}}\ \emph {et~al.}(2018)\citenamefont {{Xu}},
  \citenamefont {{Hou}}, \citenamefont {{Gong}},\ and\ \citenamefont
  {{Wang}}}]{Xu18JCAP}%
  \BibitemOpen
  \bibfield  {author} {\bibinfo {author} {\bibfnamefont {Z.}~\bibnamefont
  {{Xu}}}, \bibinfo {author} {\bibfnamefont {X.}~\bibnamefont {{Hou}}},
  \bibinfo {author} {\bibfnamefont {X.}~\bibnamefont {{Gong}}}, \ and\ \bibinfo
  {author} {\bibfnamefont {J.}~\bibnamefont {{Wang}}},\ }\href {\doibase
  10.1088/1475-7516/2018/09/038} {\bibfield  {journal} {\bibinfo  {journal} {J.
  Cosmol. Astropart. Phys.}\ }\textbf {\bibinfo {volume} {2018}},\ \bibinfo
  {eid} {038} (\bibinfo {year} {2018})},\ \Eprint
  {http://arxiv.org/abs/1803.00767} {arXiv:1803.00767 [gr-qc]} \BibitemShut
  {NoStop}%
\bibitem [{\citenamefont {{Mo}}\ \emph {et~al.}(2010)\citenamefont {{Mo}},
  \citenamefont {{van den Bosch}},\ and\ \citenamefont {{White}}}]{Mo10book}%
  \BibitemOpen
  \bibfield  {author} {\bibinfo {author} {\bibfnamefont {H.}~\bibnamefont
  {{Mo}}}, \bibinfo {author} {\bibfnamefont {F.~C.}\ \bibnamefont {{van den
  Bosch}}}, \ and\ \bibinfo {author} {\bibfnamefont {S.}~\bibnamefont
  {{White}}},\ }\href@noop {} {\emph {\bibinfo {title} {{Galaxy Formation and
  Evolution}}}}\ (\bibinfo  {publisher} {Cambridge University Press},\ \bibinfo
  {address} {Cambridge},\ \bibinfo {year} {2010})\BibitemShut {NoStop}%
\bibitem [{\citenamefont {{Shakeshaft}}(1974)}]{Shakeshaft74}%
  \BibitemOpen
  \bibinfo {editor} {\bibfnamefont {J.~R.}\ \bibnamefont {{Shakeshaft}}},\
  ed.,\ \href@noop {} {\emph {\bibinfo {title} {The Formation and Dynamics of
  Galaxies}}},\ \bibinfo {series} {IAU Symposium}, Vol.~\bibinfo {volume} {58}\
  (\bibinfo {year} {1974})\BibitemShut {NoStop}%
\bibitem [{\citenamefont {{Azreg-A{\"\i}nou}}(2014)}]{Azreg-Ainou14PRD}%
  \BibitemOpen
  \bibfield  {author} {\bibinfo {author} {\bibfnamefont {M.}~\bibnamefont
  {{Azreg-A{\"\i}nou}}},\ }\href {\doibase 10.1103/PhysRevD.90.064041}
  {\bibfield  {journal} {\bibinfo  {journal} {Phys. Rev. D}\ }\textbf {\bibinfo
  {volume} {90}},\ \bibinfo {eid} {064041} (\bibinfo {year} {2014})},\ \Eprint
  {http://arxiv.org/abs/1405.2569} {arXiv:1405.2569 [gr-qc]} \BibitemShut
  {NoStop}%
\bibitem [{\citenamefont {{Hou}}\ \emph
  {et~al.}(2018{\natexlab{b}})\citenamefont {{Hou}}, \citenamefont {{Xu}},
  \citenamefont {{Zhou}},\ and\ \citenamefont {{Wang}}}]{Hou18JCAP}%
  \BibitemOpen
  \bibfield  {author} {\bibinfo {author} {\bibfnamefont {X.}~\bibnamefont
  {{Hou}}}, \bibinfo {author} {\bibfnamefont {Z.}~\bibnamefont {{Xu}}},
  \bibinfo {author} {\bibfnamefont {M.}~\bibnamefont {{Zhou}}}, \ and\ \bibinfo
  {author} {\bibfnamefont {J.}~\bibnamefont {{Wang}}},\ }\href {\doibase
  10.1088/1475-7516/2018/07/015} {\bibfield  {journal} {\bibinfo  {journal} {J.
  Cosmol. Astropart. Phys.}\ }\textbf {\bibinfo {volume} {2018}},\ \bibinfo
  {eid} {015} (\bibinfo {year} {2018}{\natexlab{b}})},\ \Eprint
  {http://arxiv.org/abs/1804.08110} {arXiv:1804.08110 [gr-qc]} \BibitemShut
  {NoStop}%
\bibitem [{\citenamefont {{Xu}}\ \emph {et~al.}(2021)\citenamefont {{Xu}},
  \citenamefont {{Wang}},\ and\ \citenamefont {{Tang}}}]{Xu21JCAP}%
  \BibitemOpen
  \bibfield  {author} {\bibinfo {author} {\bibfnamefont {Z.}~\bibnamefont
  {{Xu}}}, \bibinfo {author} {\bibfnamefont {J.}~\bibnamefont {{Wang}}}, \ and\
  \bibinfo {author} {\bibfnamefont {M.}~\bibnamefont {{Tang}}},\ }\href
  {\doibase 10.1088/1475-7516/2021/09/007} {\bibfield  {journal} {\bibinfo
  {journal} {J. Cosmol. Astropart. Phys.}\ }\textbf {\bibinfo {volume}
  {2021}},\ \bibinfo {eid} {007} (\bibinfo {year} {2021})},\ \Eprint
  {http://arxiv.org/abs/2104.13158} {arXiv:2104.13158 [gr-qc]} \BibitemShut
  {NoStop}%
\bibitem [{\citenamefont {{Yang}}\ \emph {et~al.}(2024)\citenamefont {{Yang}},
  \citenamefont {{Liu}}, \citenamefont {{{\"O}vg{\"u}n}}, \citenamefont
  {{Lambiase}},\ and\ \citenamefont {{Long}}}]{Yang24EPJC}%
  \BibitemOpen
  \bibfield  {author} {\bibinfo {author} {\bibfnamefont {Y.}~\bibnamefont
  {{Yang}}}, \bibinfo {author} {\bibfnamefont {D.}~\bibnamefont {{Liu}}},
  \bibinfo {author} {\bibfnamefont {A.}~\bibnamefont {{{\"O}vg{\"u}n}}},
  \bibinfo {author} {\bibfnamefont {G.}~\bibnamefont {{Lambiase}}}, \ and\
  \bibinfo {author} {\bibfnamefont {Z.-W.}\ \bibnamefont {{Long}}},\ }\href
  {\doibase 10.1140/epjc/s10052-024-12412-6} {\bibfield  {journal} {\bibinfo
  {journal} {Eur. Phys. J. C}\ }\textbf {\bibinfo {volume} {84}},\ \bibinfo
  {eid} {63} (\bibinfo {year} {2024})},\ \Eprint
  {http://arxiv.org/abs/2308.05544} {arXiv:2308.05544 [gr-qc]} \BibitemShut
  {NoStop}%
\bibitem [{\citenamefont {Hawking}\ and\ \citenamefont
  {Ellis}(2023)}]{Hawking:1973uf}%
  \BibitemOpen
  \bibfield  {author} {\bibinfo {author} {\bibfnamefont {S.~W.}\ \bibnamefont
  {Hawking}}\ and\ \bibinfo {author} {\bibfnamefont {G.~F.~R.}\ \bibnamefont
  {Ellis}},\ }\href {\doibase 10.1017/9781009253161} {\emph {\bibinfo {title}
  {{The Large Scale Structure of Space-Time}}}},\ Cambridge Monographs on
  Mathematical Physics\ (\bibinfo  {publisher} {Cambridge University Press},\
  \bibinfo {year} {2023})\BibitemShut {NoStop}%
\bibitem [{\citenamefont {Ghosh}\ and\ \citenamefont
  {Kothawala}(2008)}]{Ghosh:2008zza}%
  \BibitemOpen
  \bibfield  {author} {\bibinfo {author} {\bibfnamefont {S.~G.}\ \bibnamefont
  {Ghosh}}\ and\ \bibinfo {author} {\bibfnamefont {D.}~\bibnamefont
  {Kothawala}},\ }\href {\doibase 10.1007/s10714-007-0511-6} {\bibfield
  {journal} {\bibinfo  {journal} {Gen. Rel. Grav.}\ }\textbf {\bibinfo {volume}
  {40}},\ \bibinfo {pages} {9} (\bibinfo {year} {2008})},\ \Eprint
  {http://arxiv.org/abs/0801.4342} {arXiv:0801.4342 [gr-qc]} \BibitemShut
  {NoStop}%
\bibitem [{\citenamefont {Kothawala}\ and\ \citenamefont
  {Ghosh}(2004)}]{Kothawala:2004fy}%
  \BibitemOpen
  \bibfield  {author} {\bibinfo {author} {\bibfnamefont {D.}~\bibnamefont
  {Kothawala}}\ and\ \bibinfo {author} {\bibfnamefont {S.~G.}\ \bibnamefont
  {Ghosh}},\ }\href {\doibase 10.1103/PhysRevD.70.104010} {\bibfield  {journal}
  {\bibinfo  {journal} {Phys. Rev. D}\ }\textbf {\bibinfo {volume} {70}},\
  \bibinfo {pages} {104010} (\bibinfo {year} {2004})},\ \Eprint
  {http://arxiv.org/abs/1007.2500} {arXiv:1007.2500 [gr-qc]} \BibitemShut
  {NoStop}%
\bibitem [{\citenamefont {Ghosh}\ and\ \citenamefont
  {Kumar}(2020)}]{Ghosh:2020syx}%
  \BibitemOpen
  \bibfield  {author} {\bibinfo {author} {\bibfnamefont {S.~G.}\ \bibnamefont
  {Ghosh}}\ and\ \bibinfo {author} {\bibfnamefont {R.}~\bibnamefont {Kumar}},\
  }\href {\doibase 10.1088/1361-6382/abc134} {\bibfield  {journal} {\bibinfo
  {journal} {Class. Quant. Grav.}\ }\textbf {\bibinfo {volume} {37}},\ \bibinfo
  {pages} {245008} (\bibinfo {year} {2020})},\ \Eprint
  {http://arxiv.org/abs/2003.12291} {arXiv:2003.12291 [gr-qc]} \BibitemShut
  {NoStop}%
\bibitem [{\citenamefont {Toshmatov}\ \emph {et~al.}(2017)\citenamefont
  {Toshmatov}, \citenamefont {Bambi}, \citenamefont {Ahmedov}, \citenamefont
  {Abdujabbarov},\ and\ \citenamefont {Stuchl\'\i{}k}}]{Toshmatov:2017kmw}%
  \BibitemOpen
  \bibfield  {author} {\bibinfo {author} {\bibfnamefont {B.}~\bibnamefont
  {Toshmatov}}, \bibinfo {author} {\bibfnamefont {C.}~\bibnamefont {Bambi}},
  \bibinfo {author} {\bibfnamefont {B.}~\bibnamefont {Ahmedov}}, \bibinfo
  {author} {\bibfnamefont {A.}~\bibnamefont {Abdujabbarov}}, \ and\ \bibinfo
  {author} {\bibfnamefont {Z.}~\bibnamefont {Stuchl\'\i{}k}},\ }\href {\doibase
  10.1140/epjc/s10052-017-5112-2} {\bibfield  {journal} {\bibinfo  {journal}
  {Eur. Phys. J. C}\ }\textbf {\bibinfo {volume} {77}},\ \bibinfo {pages} {542}
  (\bibinfo {year} {2017})},\ \Eprint {http://arxiv.org/abs/1702.06855}
  {arXiv:1702.06855 [gr-qc]} \BibitemShut {NoStop}%
\bibitem [{\citenamefont {Misner}\ \emph {et~al.}(1973)\citenamefont {Misner},
  \citenamefont {Thorne},\ and\ \citenamefont {Wheeler}}]{Misner73}%
  \BibitemOpen
  \bibfield  {author} {\bibinfo {author} {\bibfnamefont {C.~W.}\ \bibnamefont
  {Misner}}, \bibinfo {author} {\bibfnamefont {K.~S.}\ \bibnamefont {Thorne}},
  \ and\ \bibinfo {author} {\bibfnamefont {J.~A.}\ \bibnamefont {Wheeler}},\
  }\href@noop {} {\emph {\bibinfo {title} {Gravitation}}}\ (\bibinfo
  {publisher} {W. H. Freeman},\ \bibinfo {address} {San Francisco},\ \bibinfo
  {year} {1973})\BibitemShut {NoStop}%
\bibitem [{\citenamefont {{Dadhich}}\ and\ \citenamefont
  {{Shaymatov}}(2022{\natexlab{a}})}]{Dadhich22a}%
  \BibitemOpen
  \bibfield  {author} {\bibinfo {author} {\bibfnamefont {N.}~\bibnamefont
  {{Dadhich}}}\ and\ \bibinfo {author} {\bibfnamefont {S.}~\bibnamefont
  {{Shaymatov}}},\ }\href {\doibase 10.1016/j.dark.2022.100986} {\bibfield
  {journal} {\bibinfo  {journal} {Phys. Dark Universe}\ }\textbf {\bibinfo
  {volume} {35}},\ \bibinfo {eid} {100986} (\bibinfo {year}
  {2022}{\natexlab{a}})},\ \Eprint {http://arxiv.org/abs/2104.00427}
  {arXiv:2104.00427 [gr-qc]} \BibitemShut {NoStop}%
\bibitem [{\citenamefont {{Dadhich}}\ and\ \citenamefont
  {{Shaymatov}}(2022{\natexlab{b}})}]{Dadhich22IJMPD}%
  \BibitemOpen
  \bibfield  {author} {\bibinfo {author} {\bibfnamefont {N.}~\bibnamefont
  {{Dadhich}}}\ and\ \bibinfo {author} {\bibfnamefont {S.}~\bibnamefont
  {{Shaymatov}}},\ }\href {\doibase 10.1142/S0218271821501200} {\bibfield
  {journal} {\bibinfo  {journal} {Int. J. Mod. Phys. D}\ }\textbf {\bibinfo
  {volume} {31}},\ \bibinfo {eid} {2150120} (\bibinfo {year}
  {2022}{\natexlab{b}})}\BibitemShut {NoStop}%
\end{thebibliography}%

\end{document}